\documentclass[%
 aip,
 amsmath,amssymb,
 preprint,
 floatfix
]{revtex4-1}

\usepackage{graphicx}
\usepackage{dcolumn}
\usepackage{bm}

\usepackage[utf8]{inputenc}
\usepackage[T1]{fontenc}
\usepackage{mathptmx}

\usepackage{relsize}
\usepackage{placeins}
\usepackage{booktabs}

\begin{document}

\preprint{}

\title[Time delay effects in the control of synchronous electricity grids]{Time delay effects in the control of synchronous electricity grids}

\author{Philipp C. B\"ottcher}
\email{philipp.boettcher@dlr.de}
\affiliation{%
 DLR-Institute of Networked Energy Systems, Carl-von-Ossietsky Stra{\ss}e 15, 26129 Oldenburg, Germany\\
}%

\author{Andreas Otto}
\email{otto.a@mail.de}
\affiliation{Institute of Physics, Chemnitz University of Technology, 09107 Chemnitz, Germany\\
}%

\author{Stefan Kettemann}
\affiliation{Jacobs University, Department of Physics \& Earth Sciences, Campus Ring 1, 28759 Bremen\\
}%
\affiliation{
Division of Advanced Materials Science, Pohang University of Science and Technology (POSTECH),San 31, Hyoja-dong, Nam-gu, Pohang 790-784, South Korea}
\author{Carsten Agert}
\affiliation{%
 DLR-Institute of Networked Energy Systems, Carl-von-Ossietsky Stra{\ss}e 15, 26129 Oldenburg, Germany\\
}%

\date{\today}

\begin{abstract}
The expansion of inverter-connected generation facilities (i.e. wind and photovoltaics) and the removal of conventional power plants is necessary to mitigate the impacts of climate change.
Whereas conventional generation with large rotating generator masses provides stabilizing inertia, inverter-connected generation does not.
Since the underlying power system and the control mechanisms that keep it close to a desired reference state, were not designed for such a low inertia system, this might make the system vulnerable to disturbances.
In this paper, we will investigate whether the currently used control mechanisms are able to keep a low inertia system stable and how this is effected by the time delay between a frequency deviation and the onset of the control action.
We integrate the control mechanisms used in continental Europe into a model of coupled oscillators which resembles the second order Kuramoto model.
This model is then used to investigate how the interplay of changing inertia, network topology and delayed control effects the stability of the interconnected power system.
To identify regions in parameter space that make stable grid operation possible, the linearized system is analyzed to create the system's stability chart.
We show that lower and distributed inertia could have a beneficial effect on the stability of the desired synchronous state.
\end{abstract}

\maketitle

\begin{quotation}
Reducing the share of fossil fuel based power generation is a key factor in fighting climate change.
To maintain the overall energy generation, they need to be replaced by generation from renewable resources.
The currently used control mechanisms to ensure a stable electric power system have been established upon the experience with so-called conventional energy resources. 
Thus it is necessary to examine if the currently used control mechanisms can cope with this transition to a power system dominated by renewable generation.
In order to achieve this, we include these control mechanisms in a model describing the dynamics of the interconnected power system and take into account their delayed reaction.
Our findings suggest that reducing the amount of conventional generation by introducing a higher share of renewable generation and distributing the renewable generation throughout the system, makes the system more stable in case of time delays in the control mechanism. 
\end{quotation}


\section{\label{sec:intro}Introduction}

The transition towards a power system that relies on renewable resources presents a major challenge to the energy system\cite{sims2004renewable}.
During the transition, highly volatile energy sources (i.e. wind and photovoltaics) will be introduced\cite{iwes2015european} to a system built with conventional energy sources in mind.
Presently, the power frequency control operated by the 'European Network of Transmission System Operators for Electricity' (ENTSO-E) guarantees the stable of operation of this interconnected system.
These control mechanisms can only be employed by accurately measuring the system state (i.e. frequencies and load flows) and by correctly communicating theses values.
At present, conventional generation (e.g. thermal power plants) with large rotating generator masses provide stabilizing inertia to the system.
Removing these conventional generation facilities and replacing them with fluctuating renewable generation that does not provide inertia could make the system vulnerable to disturbances and accelerate dynamics\cite{wu2015preliminary, tielens2016relevance}.
The delay associated with the measurement, communication and the deployment of control might play an increasingly important role in a system that relies on inertia-less feed-in that is fluctuating on small time scales.\\
In the context of complex systems research, the stability and dynamics of power grids have been studied.
One approach is to consider energy systems or more specifically power grids as complex networks of coupled oscillators described by Kuramoto-like models \cite{filatrella2008analysis, rohden2012self, witthaut2012braess, rohden2014impact,witthaut2016critical, tchuisseu2018curing}. 
The main goal of related studies is to identify the limits of synchronous operation of the power transmission network. 
The collective frequency is not, as one might suspect, the average of the frequencies of the individual nodes. 
Instead they are related to the topology, i.e. the contributions of the individual oscillators are weighted with their centrality in the network \cite{Skardal2016}.
The examination of the transmission network itself can reveal certain weaknesses of the network and help to guarantee a robust and stable system. 
Witthaut et al. \cite{witthaut2016critical} showed that critical links are not only determined by their typical load but also by features of the network's global topology.
Thus, effects that emerge in transport networks, e.g. Braess's paradox, have been shown to be present in power grids \cite{witthaut2012braess, tchuisseu2018curing}.
According to Rohden et al. \cite{rohden2012self, rohden2014impact}, a higher share of decentralized energy production promotes the structural robustness of the resulting energy system but makes the system more susceptible to short-term perturbations, necessitating rigorous control mechanisms and an understanding of how to distribute inertia thorough out the system \cite{jacquod2019optimal}. How does distributed inertia affect the system stability in the presence of delayed control?\\
In the context of power grids, delay has been shown to have a destabilizing effect on the dynamics of power grids modelled as networks of coupled phase oscillators \cite{schafer2015decentral, schafer2016taming}.
Even time averaging over past states can not guarantee a stable system.
In general, systems with delay, also called time delay systems, can be described by delay differential equations (DDEs). It is known that delays can have both stabilizing as well as de-stabilizing effects \cite{otto2019delaydynamics, sipahi2011ddestability}. 
In DDEs the stability of a fixed point can switch from stable to unstable and back again multiple times under variation of the delay \cite{sipahi2002ddeexact, lakshmanan2011dynamics}.
With the knowledge of regions in parameter space where the fixed point is stable, the stability can be enhanced by tuning the parameters or the delay \cite{bokharaie2014small}.\\
In this paper, the load frequency control that is currently being used in Europe\cite{machowskipower, handbook2009policy} is incorporated into the model of coupled oscillators by taking into account the two fastest automatic control mechanisms\cite{andersson2012dynamics} (i.e. primary and secondary control).
We consider a Kuramoto-like model of the electricity grid, where each oscillator corresponds to one control area, and we introduce a time delay into the feedback control mechanisms of each control area.
While we neglect the effects of time delay in the primary control, a time delay is introduced in the slower secondary control.\\
In particular, we discuss the basic concepts and general trends by considering  a simple system consisting of two control areas.
Subsequently, two larger control area networks will be examined:
a tree-like network commonly known as the Cayley tree, and a system that more closely resembles the control area network of continental Europe which was extracted from open data.\\
The main objective of this work is to investigate the stability behavior of the equilibria related to stable grid operation and to present stability charts to show the effects of different changes to the control area network (e.g. different inertia, control gains and different network topologies) on the stability of the power grid when considering delayed control.
For each of the considered networks, cases with homogeneously and inhomogeneously distributed inertia will be compared.
The general trend in all of these examples shows that one can in principle increase the stability of the desired operating state by decreasing and distributing inertia intelligently.\\ 
The paper is organized as follows.
The power grid model with the considered control mechanisms is introduced in Sec.~\ref{sec:model}. 
In Sec.~\ref{sec:theory}, we present the linear stability analysis and the numerical methods for constructing the stability charts.
Results concerning the stability behavior of the different control area networks under varying parameters can be found in Sec. \ref{sec:results}.
The main results and implications for the power system are summarised in Sec.~\ref{sec:conclusion}.

\section{\label{sec:model} Modelling the Frequency Dynamics}

\subsection{Power Grid Model}
The European power system consists of many different components e.g. generating units, loads and transmission lines. 
These are connected at different voltage levels.
A distinction is made between the network used to deliver power over large distances and the system designed to supply end consumers with electricity.
They are referred to as transmission system and the distribution system, governed by the transmission system operators (TSOs) and the distribution system operators, respectively.\\
As we examine the frequency dynamics of the interconnected power system, we consider only the highest grid level, i.e. the transmission system.
This is reasonable since frequency dynamics is mainly subject to the large scale interaction of the entire power system, while the voltage dynamics are subject to local phenomena.\\
The control mechanisms that keep the frequency close to the reference frequency (i.e. 50Hz in Europe) are defined on the level of TSOs, which are together responsible for the load-frequency control in Europe. 
To achieve this they are organized in the ENTSO-E, which governs the rules and regulations that are needed to cooperatively keep the system stable  \cite{handbook2004policy, handbook2004appendix, handbook2009policy}.
The ENTSO-E splits Europe into regional groups with Continental Europe being the largest one.
These regions are further split into control areas that run synchronously to each other with a nominal frequency of $\omega_0 = 2\pi \; 50$Hz. 
TSOs are responsible for the load-frequency control in their respective control area.\\
In this paper, we consider $N$ control areas, where each area is modeled as one aggregated machine\cite{kundur1994power, ulbig2014impact}.
Analogous to a synchronous machine, this aggregated machine $i$ is characterized by a power phase angle $\phi_i=\omega_0 \; t+\theta_i$, where $\theta_i$ denotes the deviations from the nominal phase angle $\omega_0 \; t$.
Using the model for a network of synchronous machines for high voltage transmission grids described in Ref.~\cite{filatrella2008analysis}, the dynamics of the power phase angle $\theta_i$ of area $i$ is given by  
\begin{equation}
    A_i \ddot{\theta}_i(t) + k_{l,i} \dot{\theta}_i(t) + \sum_{j=1}^N C_{ij} \sin{(\theta_i(t) - \theta_j(t))} = P_{i,0}+P_{\text{c},i}(t) \label{eq:norm_swing}
\end{equation}
where we have used $A_i = 2 H_i\; S_{B,i}/{\omega_0}$.
The coupling via transmitted power is governed by the so-called power flow equations derived from Kirchhoff's laws\cite{machowskipower}.
Since we are only concerned with the transmission system consisting of the highest voltage levels, we assume a lossless, purely inductive transmission of power.
In this case, only active power needs to be considered given by the transmission capacity $C_{ij}$ between area $i$ and $j$ and the sine of the power phase angle differences.
$A_i$ is product of the share of inertia providing generation quantified by the inertia constant $H_i$ and the size of area $i$ in terms of power $S_{B, i}$.
Thus, it is proportional to the total inertia that area $i$ provides.
$H_i$ is a measure of how long the rated power $S_{B,i}$ can be supplied by the kinetic energy of the rotating generator masses.
A low $H_i$ indicates a situation with a high share of the produced electricity in area $i$ by inverter-connected generation.\\
In addition to a constant loss due to dissipation, frequency-dependent load damping occurs for larger power systems.
This effect, commonly known as self-regulation \cite{kurth2006importance}, summarizes the present time-varying dissipation effects and is given by $P_{\text{diss},i}(t)=k_{l,i} \dot{\theta}_i(t)$, where $k_{l,i}=k_l S_{B,i}$ and $k_l$ gives the fraction of load that is assumed to contribute to this effect ($k_l\approx 1\%/$Hz).\\
Eq.~\eqref{eq:norm_swing} closely resembles the second order Kuramoto model with inertia, which is a prototypical model for synchronization in complex networks\cite{rodrigues2016kuramoto}.
The existence of a synchronized state with a common frequency $\dot{\theta}_i(t)=\omega_i=\omega \; \forall \; i$, in our case the synchronous operation with $\omega_i=0$, can be observed for sufficiently high transmission capacities $C_{ij}$ \cite{rohden2012self}.\\
In this paper, we consider only networks, where this synchronous state exists.
In the ideal synchronous grid operation with $\omega_i=0 \; \forall \; i$ and stationary phases $\theta_i(t)=\theta_{i,0}$, stationary power flows remain which are given by the distribution of the stationary power injections $P_{i,0}$.
While a coexistence of limit cycles and the fixed point of synchronous operation may be observed \cite{rohden2012self}, we will focus on the fixed point corresponding to synchronous operation for balanced areas (i.e. $P_{i,0}=0$) and how its stability is affected by the delayed control power $P_{\text{c},i}(t)$.

\subsection{\label{sec:eu_sys} Control of the European Power System}
\begin{figure}
    \centering
    \includegraphics[width=\columnwidth]{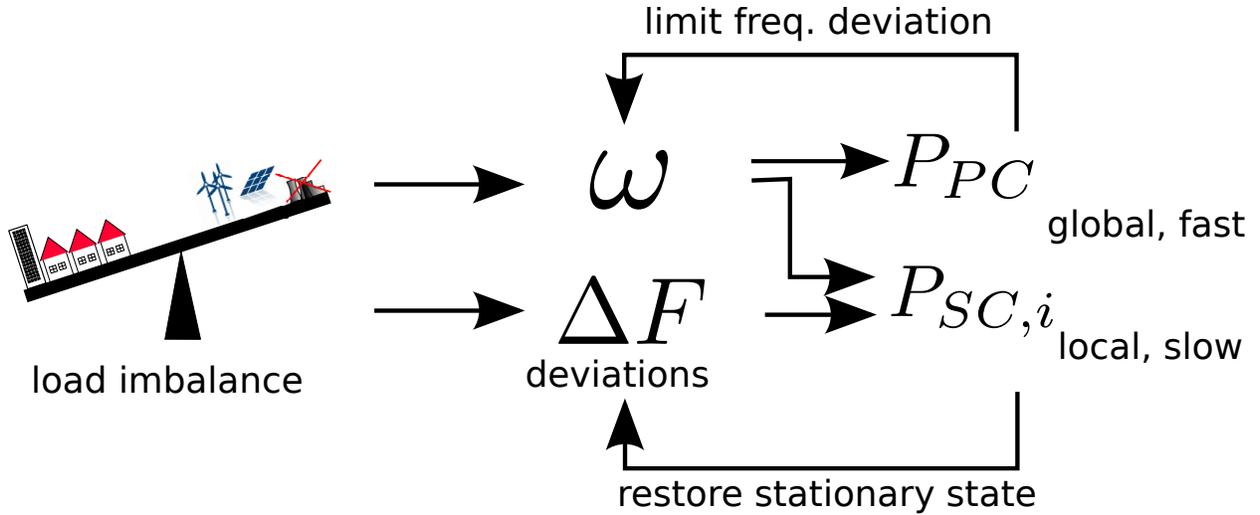}
    \caption{Load-frequency control scheme with primary and secondary control, which are considered in this paper.
    Imbalances in generation and consumption in control area $i$ lead to deviations of the frequency $\omega_i$ and the power flow $\Delta F_i$ to neighbouring areas.
    Primary control $P_{\text{PC}}$ counteracts the imbalance of production and consumption within seconds of a detected disturbance and limits the frequency deviation. Secondary control $P_{\text{SC}, i}$ brings the frequency back to the reference value and restores the predisturbance state.
    It is activated within a few seconds and remains active for up to 15 minutes. }
    \label{fig:control_scheme}
\end{figure}

An important quality factor in synchronous electricity grids is the grid frequency.
Its nominal value $\omega_0$ (50 Hz in Europe) is chosen by keeping different factors like losses and costs in mind \cite{schavemaker2017electrical}.
Imbalances in supply and demand of power lead to deviations from $\omega_0$.
For example, if a power plant is disconnected from the grid by some contingency, the grid frequency changes to a lower value.
The rate of change is determined by the inertia.
Inertia is provided mainly by large rotating generator masses in conventional generation facilities.
The amount of inertia that effects the frequency dynamics is not constant.
It depends on the share of currently connected inertia providing (e.g. conventional generation) and inverter-connected (e.g. solar or wind) generation\cite{ulbig2014impact}.\\
Disturbances, that lead to a frequency deviation, propagate through the system.
The behavior of this propagation varies depending on the system's parameters and the nature of the disturbance. 
For low inertia this can lead to a delocalization of the disturbance\cite{kettemann2016delocalization, tamrakar2018propagation}.
Additionally, the fluctuations fed to the grid by renewable generation (e.g. by wind or solar) influence the grid frequency dynamics.
For example, turbulent wind fluctuations become noticeable at times with high feed-in ratios of wind power \cite{haehne2018footprint}.
These fluctuations are more pronounced in regions where a lot of power is injected by wind turbines \cite{haehne2019propagation}, which is even more pronounced when considering heterogeneities in the parameters \cite{wolff2019heterogeneities}.\\
Since frequencies outside a certain band around $\omega_0$ put devices in danger, control of the grid frequency has to be employed.
A sufficiently high back-up of control power is an ancillary service provided by power plants in addition to the generation they deliver to match the expected load.
There are different control mechanisms, which act on different time scales and serve different purposes.
Here, we consider the two fastest control mechanisms that operate automatically, namely primary ($P_{\text{PC}}$) and secondary control ($P_{\text{SC}}$).
\begin{equation}
    P_{\text{c},i}(t) =P_{\text{PC},i}(t) + P_{\text{SC},i}(t).
    \label{eq:control}
\end{equation}
Their interaction while clearing an imbalance in generation and consumption is visualized in Fig.~\ref{fig:control_scheme}.\\ 
The fastest one is primary control, which is activated within the first seconds after a disturbance has been detected.
After $30$s the full primary control power $P_{\text{PC},i}(t) = -\lambda_i \omega_i$ has to be delivered according to the guidelines of the ENTSO-E\cite{handbook2009policy}.
Its sensitivity to the frequency deviation $\omega_i$ is given by the network-power frequency characteristic $\lambda_i$ for area $i$.
It specifies the characteristic power disturbance for a given frequency deviation, is measured regularly, and kept constant for some time.
The magnitude of $\lambda_i$ depends on the makeup of the examined system and its sum $\lambda_{\text{total}}$ is measured empirically \cite{asal1998development}.
If the power disturbance is counteracted by $P_{\text{PC},i}$, the frequency does not change anymore but the system now operates at a different frequency than $\omega_0$.\\
Secondary control is used to restore the pre-disturbance configuration, specified by rotations at the nominal grid frequency $\omega_0$ ($\omega_i=0$, $\theta_i(t)=\theta_{i,0}$). The magnitude of secondary control is given by a proportional integral (PI) controller and it is used to correct the local area-control-error $G_i$\cite{handbook2004appendix}.
The power $P_{\text{SC}, i}(t)$ that is provided by secondary control is determined by
\begin{equation}
    P_{\text{SC}, i}(t) = - \left( K_P G_i(t-\tau) + K_I \int\limits_{-\infty}^{t-\tau} G_i(t') dt' \right), \label{eq:SC}
\end{equation}
with $K_P$ and $K_I$ being the tunable gain factors of the proportional and integral term, respectively.
The time delay $\tau$ specifies the time that is required for the determination of the local area-control error $G_i$, communication and the initiation of a control action.
$G_i$ is a measure of the power that is missing in area $i$.
It is determined by the difference between the expected primary control power and the deviations $\Delta F_i$ of the power flows to neighbouring control areas
\begin{align}
    G_i &= \lambda_i \omega_i - \Delta F_{i}, \\
    \Delta F_i &= \sum_j C_{ij} \left[ \sin{(\theta_{j}(t)-\theta_i(t))} - \sin{(\theta_{j,0}-\theta_{i,0})} \right]
\end{align}
Note, that the PI controller is linear but the local area-control error $G_i$ depends nonlinearly on the system state.
We neglect other nonlinearities (e.g. dead-band of primary control) or more specific models for power plants.
The cycle time of secondary control is required to be between 2 to 5 seconds\cite{handbook2009policy}.
While the value of delay is sure to be slightly different for different control areas and also time dependent, as a simplification we consider a constant delay $\tau$. 

\section{\label{sec:theory} Stability Analysis}
In the following section, we describe the theory for the linear stability analysis of the DDE around the desired reference state. We present an efficient frequency domain method for the calculation of the stability boundaries as well as a numerical method for the calculation of the dominant eigenvalues and the corresponding eigenvectors of the time delay system via Chebyshev discretization. 

\subsection{\label{sec:linsys}Linearized Dynamics}
We now determine the linearized dynamics around the desired reference state of the power grid.
For brevity, the time dependence is dropped and the delayed variables are given by the subscript $\tau$ ($\alpha(t-\tau) = \alpha_\tau$).
The reference state corresponding to the synchronous operation is given by the fixed point with $\omega_i(t)=0$ and $\theta_i(t)=\theta_{i,0} \forall i$. 
We consider small deviations $\alpha_i (t)=\theta_i(t) -\theta_{i,0}$ around this reference state.
With the relevant control terms Eq.~\eqref{eq:control} and Eq.~\eqref{eq:SC}, Eq.~\eqref{eq:norm_swing} can be written in terms of the deviations $\alpha_i$ as
\begin{flalign}
    A_i \Ddot{\alpha}_i +k_{l,i} \Dot{\alpha}_i &+ \lambda_i \Dot{\alpha}_i
    + \sum_j C_{ij} \sin{(\Delta {\theta_{ij}^0} + \Delta \alpha_{ij})} \nonumber \\
    + K_{P}   G_i(t-\tau)
    &+  K_I  \int_{-\infty}^{t-\tau} G_i(t') dt'= P_i^0 , 
    \label{eqn:full_eqn}
\end{flalign}
where $\Delta \theta_{ij}^0=\theta_{j,0}-\theta_{i,0}$, $\Delta \alpha_{ij}=\alpha_{j}-\alpha_{i}$, and
\begin{equation}
\begin{split}
    G_i(t) &=  \lambda_i \Dot{\alpha}_{i}(t)\\
    &+ \sum_j C_{ij} \left[ \sin{(\Delta{\theta_{ij}^0} + \Delta\alpha_{ij}(t))} - \sin{(\Delta {\theta_{ij}^0})} \right].
\end{split}
\end{equation}
From Eq.~\eqref{eqn:full_eqn} one can see that the simplified primary control just increases the system's damping and we can introduce the effective damping $c_i=k_{l} \cdot S_{B,i}+\lambda_i$. 
After linearization of Eq.~\eqref{eqn:full_eqn} around the reference solution, the linearized system is governed by
\begin{equation}
\begin{split}
    &A_i \Ddot{\alpha}_i +c_i \Dot{\alpha}_i + \sum_j l_{ij} \Delta \alpha_{ij} 
    + K_{P} \lambda_i  \dot{\alpha}_{i,\tau} + K_{P} \sum_j l_{ij} \Delta \alpha_{ij,\tau} \\
    &+ K_I  \int_{-\infty}^{t-\tau} \left(\lambda_i \dot{\alpha}_i+\sum_j l_{ij} \Delta \alpha_{ij} \right) dt'= 0.
    \label{eqn:full_linear_eqn}
\end{split}
\end{equation}
In the linearized system the coupling between the nodes (i.e. control areas) is described by the elements $l_{ij}$ of the weighted Laplacian ${\bf L}$, which are given by
\begin{align}
l_{ij} = \begin{cases} -C_{ij} \cdot \cos{(\theta_i^0 - \theta_j^0)} &\mbox{if } i \neq j \\
-\sum_{i\neq j}l_{ij} & \mbox{if } i = j \end{cases}.
\end{align}
The linearized system has been extensively studied in the analysis of power system stability \cite{rohden2012self, rohden2014impact}, transient dynamics and propagation of disturbances in power grids\cite{kettemann2016delocalization,zhang2019fluctuation, tyloo2018robustness}.
Note that the stationary power input $P_{i,0}$ is missing in Eq.~\eqref{eqn:full_linear_eqn} because it is equivalent to the sum $\sum\nolimits_j C_{ij} \cdot \sin{(\Delta \theta_{ij}^0)}$, and was subtracted from both sides of the equation.\\
Eq.~\eqref{eqn:full_linear_eqn} describes the dynamics of the deviations in the $i$th area of the network.
The deviations of the whole power grid at the time $t$ can be summarized in the $3N$ dimensional vector  
\begin{equation}
\begin{split}
\vec{x}(t) = \left[ \int\nolimits_{-\infty}^{t} \alpha_1(t') dt', \ldots, \int\nolimits_{-\infty}^{t} \alpha_N(t') dt', \right.\\
\left. \alpha_1(t) \ldots \alpha_N(t), \dot{\alpha}_1(t),\ldots,\dot{\alpha}_N(t)\vphantom{\int\nolimits_{1}^2}\right]^T, 
\end{split}
\end{equation} 
and its dynamics can be described in first-order form as
\begin{align}
\label{eq:full_lin_dyn}
    \mathbf{A} \dot{\vec{x}}(t) = \mathbf{N} \vec{x}(t) + \mathbf{D} \vec{x}(t-\tau).
\end{align}
The matrix $\mathbf{A}$ is a diagonal matrix, where the first $2N$ diagonal elements are one and the last $N$ diagonal elements are equal to $A_i$ ($A_{2N+i,2N+i} = A_i$ for $i=1,\ldots,N$).
The coefficient matrix $\mathbf{N}$ for the non-delayed term is a block-matrix given by
\[ \mathbf{N} =  \left( \begin{array}{ccc}
\mathbf{0} & \mathbf{I} & \mathbf{0}\\
\mathbf{0} & \mathbf{0} & \mathbf{I} \\
\mathbf{0} & -\mathbf{L} & -\mathbf{B}
\end{array} \right), 
\]
where $\mathbf{0}$ and $\mathbf{I}$ are the $N$ dimensional quadratic null matrix, and the identity matrix, respectively.
$\textbf{B}$ is an $N$ dimensional diagonal matrix with the damping values $c_i$ on its main diagonal ($B_{ii}=c_i$).
The coefficient matrix $\mathbf{D}$ of the delay term contains the proportional and the integral term of the delayed secondary control, and can be determined by $\mathbf{D}=-K_P \mathbf{D}_P- K_I\mathbf{D}_I$ with
\[
\mathbf{D}_P =
\left( \begin{array}{ccc}
\mathbf{0} & \mathbf{0} & \mathbf{0}\\
\mathbf{0} & \mathbf{0} & \mathbf{0}\\
\mathbf{0} & \mathbf{L}& \Lambda
\end{array} \right), \text{ and }
\mathbf{D}_I =
\left( \begin{array}{ccc}
\mathbf{0} & \mathbf{0} & \mathbf{0}\\
\mathbf{0} & \mathbf{0} & \mathbf{0}\\
\mathbf{L}& \Lambda & \mathbf{0}
\end{array} \right). 
\]
Here, $\Lambda$ is an $N$ dimensional diagonal matrix with the coefficients for primary control $\lambda_i$ on its diagonal ($\Lambda_{ii}=\lambda_i$).

\subsection{\label{sec:theory_stablobes}Stability Boundaries}
Eq.~\eqref{eq:full_lin_dyn} is a linear DDE with constant coefficients.
Linear DDEs have eigenmodes of the form $\vec{x}(t)=\vec{x}(0) \frac{1}{2}\left(e^{s t}+e^{s^* t}\right)$ (see ref. \cite{amann2007some}), where $s \in \mathbb{C}$ are called characteristic roots and $s^*$ denotes the complex conjugate of $s$.
The characteristic roots are the roots of the characteristic equation, which can be obtained by putting the exponential ansatz $\vec{x}=\vec{v}e^{s t}$ in the DDE.
The characteristic equation for the DDE Eq.~\eqref{eq:full_lin_dyn} is given by
\begin{equation}
    \det\left(\mathbf{A}s - \mathbf{N} + \left( K_P \mathbf{D}_P + K_I \mathbf{D}_I \right)e^{-s \tau}\right) = 0.
    \label{eq:char_eqn}
\end{equation}
Due to the presence of the delay term, Eq.~\eqref{eq:char_eqn} is a transcendental equation and has infinitely many solutions, which means that the delay system is infinite dimensional and has infinitely many eigenmodes.
The system is stable if all characteristic roots have negative real part \cite{michiels2014ddestability}. 

We are interested in the stability boundaries given by a set of parameters values $H_i,K_P, K_I, \tau$ at which the dominant characteristic root, i.e. the characteristic root with the largest real part, crosses the imaginary axis.
In particular, we will explore the change of the stability boundaries in dependence of parameter changes. There are various methods for calculating the characteristic roots of linear time-invariant DDEs and determine its stability \cite{michiels2014ddestability, jarlebring2008}.
However, since we have three variables per node and the number of nodes $N$ in the network can become large, the system dimension can be quite large and we are interested in an efficient method for the calculation of the stability boundaries.
Such a method exists for the analysis of machine tool dynamics, where similar systems appear \cite{altintas2004chatter, otto2014extension}.
In this field the stability boundaries are called stability lobes and its calculation is important for guaranteeing stable cutting processes without undesired large vibrations.
While we use the term \textit{lobes}, which is more common in the engineering literature, the term \textit{leaves} is used in the chaos control community \cite{balanov2005delayed}. 
Here, we briefly describe a very efficient method adapted for the calculation of the limiting $K_P$ or $K_I$ in dependence of the delay $\tau$, which is described in Ref.~\cite{otto2014extension}.

The characteristic Eq.~\eqref{eq:char_eqn} can be also written as an eigenvalue equation as
\begin{equation}
\label{eq:eigenvalue_eqn_3N}
    \left(\mathbf{A}s - \mathbf{N} + \left( K_P \mathbf{D}_P + K_I \mathbf{D}_I \right)e^{-s \tau}\right)\vec{v}(s) = 0.
\end{equation}
From the structure of the system it follows that $\vec{v}(s)=\left[\vec{u}(s),s \vec{u}(s), s^2 \vec{u}(s) \right]^T$, that is, the $N$ dimensional vector $s \vec{u}(s)$ specifies for example the angular deviations of the grid in the Laplace domain.
As a consequence, the $3N$ dimensional Eq.~\eqref{eq:eigenvalue_eqn_3N} corresponding to the first-order representation is equivalent to an $N$ dimensional equation, with higher order terms in $s$. The equivalent $N$ dimensional representation can be given by
\begin{equation}
\label{eq:eigenvalue_eqn}
\left(s^3\hat{\mathbf{A}}+ s^2\mathbf{B} + s\mathbf{L}\right)e^{s \tau}\vec{u}(s)=-(s K_P  + K_I) \left(s \Lambda +\mathbf{L}\right)\vec{u}(s),
\end{equation}
where $\hat{\mathbf{A}}=diag(A_1,\dots,A_N)$ with $A_i \propto H_i \cdot S_{B,i}$ encodes the inertia and is the lower right $N \times N$ block of the matrix $\mathbf{A}$.\\
By assuming that the matrix
\begin{equation}
\mathbf{M}(s)=\left(s \Lambda +\mathbf{L}\right)^{-1} \left(s^3\hat{\mathbf{A}}+ s^2\mathbf{B} + s\mathbf{L}\right)
\end{equation}
is diagonalizable, we can substitute the eigenvalues $\sigma(s) \in \mathbb{C}$, of the matrix $\mathbf{M}(s)$ in Eq.~\eqref{eq:eigenvalue_eqn} and obtain the scalar equation
\begin{equation}
\label{eq:eigenvalue_eqn_scalar}
\sigma(s)e^{s \tau}=-(s K_P  + K_I).
\end{equation}
Eq.~\eqref{eq:eigenvalue_eqn_scalar} is another form of the characteristic equation and can be used for the calculation of the characteristic roots.
Since we have a set of scalar equations with isolated dependencies on the parameters $\tau$, $K_P$, and $K_I$,  Eq.~\eqref{eq:eigenvalue_eqn_scalar} is suitable for calculating the limiting stability boundaries in a parameter space spanned by $\tau$, $K_P$, and $K_I$.\\
The latter approach can be explained as follows. At the stability boundaries we have $s=j \eta$ with $j=\sqrt{-1}$ as the imaginary unit, i.e., the real part of the dominant characteristic root $s$ vanishes.
After substituting $s=j \eta$  in Eq.~\eqref{eq:eigenvalue_eqn_scalar} and rearranging, we obtain 
\begin{equation}
\label{eq:calc-stability-lobes}
\begin{split}
K_P &=-\frac{K_I+\sigma(j \eta)e^{j \eta \tau}}{j \eta},  \text{ or } \\
K_I &=-j \eta K_P+\sigma(j \eta)e^{j \eta \tau},
\end{split}
\end{equation}
depending on whether we would like to calculate the limiting $K_P$ or $K_I$, respectively.\\
In general, for an arbitrary imaginary part $\eta$, the right hand side of Eq.~\eqref{eq:calc-stability-lobes} is  a complex value, whereas the parameters $K_P$ and $K_I$ are real values.
Thus, by setting the imaginary part of the right hand side of Eq.~\eqref{eq:calc-stability-lobes} equal to zero, we find critical $\eta_c$'s for which one characteristic root crosses the imaginary axis.
In particular, the $\eta_c$ is the critical frequency that characterizes the dynamics close to the bifurcation point. Then, the critical gain values $K_P$ or $K_I$ can be determined by substituting $\eta=\eta_c$ in Eq.~\eqref{eq:calc-stability-lobes}.
In practice, the critical $\eta_c$ can be found by a parametric sweep of $\eta$, and comparison of the imaginary part of the right hand side of Eq.~\eqref{eq:calc-stability-lobes} for two subsequent values $\eta_k$ and $\eta_{k+1}$ of the frequency $\eta$.
Finding a solution $\eta_c$ for a given $\tau$ also gives solutions with 
\begin{align}
    \tau'= \tau + \frac{2\pi \;n}{\eta_c} \quad \text{where} \quad n \in \mathbb{Z}
    \label{eq:periodicity}
\end{align}
due to the periodicity of the $e^{j \eta \tau}$ term.
This could in principle be used to evaluate just the first lobes and continue them according to Eq.~\eqref{eq:periodicity}, thus reducing computational complexity \footnote{The authors thank Eckehard Schöll for hints in this direction.}.\\ 
For a correct identification of a zero-crossing of the imaginary part the correct mapping between the eigenvalues $\sigma(j\eta_k)$ and $\sigma(j\eta_{k+1})$ is important. Assuming that the step width $\eta_{k+1}-\eta_{k}$ is small, the eigenvector belonging to a eigenvalue does not change much for one step. This property can be used to identify corresponding eigenvalues at subsequent frequency steps by comparing their eigenvectors via the modal assurance criterion (MAC) value as described in Ref. \cite{loeser2018uncertain}.\\
The steps for the calculation of the stability lobes can be summarized as follows: 
\begin{enumerate}
    \item Specify the system parameters (i.e. $H_i, S_{B,i}, k_{l,i}, C_{ij}$ and $\lambda$ for all control areas $i$), the delay $\tau$, and $K_P$ or $K_I$.
    \item Calculate the eigenvalues $\sigma(j\eta)$ of the matrix $M(j\eta)$ for a grid of values $\eta=\eta_k$.
    \item Sort the eigenvalues $\sigma(j\eta_k)$ according to its eigenvector via the MAC value \cite{loeser2018uncertain}
    \item Find the critical frequencies $\eta_c$ for which the imaginary part of the right hand side of Eq.~\eqref{eq:calc-stability-lobes} vanishes.
    \item Calculate the critical $K_{P}$ or $K_{I}$ by substituting the critical frequencies $\eta_c$ in Eq.~\eqref{eq:calc-stability-lobes}.
\end{enumerate}
The resulting critical curves represent all parameter combinations, where at least one characteristic root $s$ of the DDE has vanishing real part.
For the stability boundaries, however, only the crossings of the dominant roots are relevant.
Since in most cases the linearized system Eq.~\eqref{eqn:full_linear_eqn} is marginally stable for $K_P=0$ or $K_I=0$, the curve at the lowest critical $K_P$ or $K_I$, respectively, represents the stability boundary that separates stable from unstable behavior (cf. Fig.~\ref{fig:res_twoarea_time_cheb}) for a given $\tau$.
While the proposed method is sufficient for the examined networks, larger networks might benefit from more sophisticated methods to determine the stability boundaries \cite{ramirez2019approach}.
In principle, isolated regions in parameter space may exist, where the fixed point is stable.
However, for the necessary conservative choice of the control gains $K_P$ and $K_I$ in applications the stability islands are of less practical interest and will not be considered here.

\subsection{\label{sec:theory_chebyshev}Computation of Dominant Roots}
Whereas the method in Sec.~\ref{sec:theory_stablobes} can be used for a very efficient calculation of the stability boundaries, it does not give any information about the eigenvalue spectrum or the corresponding eigenvectors.\\
For this purpose, we use the Chebyshev collocation method for the calculation of the dominant characteristic roots of the linear DDE Eq.~\eqref{eqn:full_linear_eqn} \cite{breda2006chebyshev, jarlebring2008}.
The reference state is not stable if any characteristic roots has a positive real part.\\ 
The Chebyshev collocation method can be described as follows. The state of the DDE Eq.~\eqref{eq:full_lin_dyn} is the function $\vec{x}(\theta)$ in the interval $[t-\tau,t]$. The state interval is discretized by using the Chebyshev points $t_k=\cos{\frac{k}{M}\pi} \in [-1,1]$, with $k=0, \ldots, M$ \cite{trefethen2000spectral}. In particular, the approximated state of the DDE can be given by the vector 
$\vec{y}(t) = \left[\vec{x}_0(t), \ldots, \vec{x}_M(t)\right]^T$, where $\vec{x}_k(t)=\vec{x}(t-\frac{\tau}{2}(t_k+1))$.
By using the $3N(M+1)$ dimensional state vector $\vec{y}(t)$ instead of the $3N$ dimensional configuration $\vec{x}(t)$, the DDE Eq.~\eqref{eq:full_lin_dyn} can be approximated via an ODE as
\begin{equation}
    \Dot{\vec{y}}(t) = \mathbf{M}_\text{C}\vec{y}(t).
\end{equation} 
The coefficient matrix is given by \cite{jarlebring2008}
\newcommand*{\temp}{\multicolumn{1}{c|}{0}}
\[\arraycolsep=1.6pt\def\arraystretch{1.5}
\mathbf{M_C} =
\left(\begin{array}{c}
- \mathlarger{\frac{2 \; \mathbf{C_{M}}}{\tau} \otimes \mathbf{I_{3N}}} \\ 
\hline
\begin{array}{ccccc}
\mathbf{A^{-1}}\mathbf{D},  & \mathbf{0} & \dots & \mathbf{0} ,& \mathbf{A^{-1}}\mathbf{N}\\
\end{array}  \end{array}\right),
\]     
where $\mathbf{C_M}$ is the Chebyshev differentiation matrix \cite{trefethen2000spectral} with the last row being deleted, $\mathbf{I_{3N}}$ is the $3N$ dimensional identity matrix and $\otimes$ denotes the Kronecker product.
The last row in the matrix $\mathbf{M_C}$ represents the original DDE Eq.~\eqref{eqn:full_linear_eqn}.
The other rows are a spectral approximation of the time derivative at the Chebyshev nodes.
The eigenvalues of the matrix $\mathbf{M_C}$ approximate the dominant characteristic roots $s$ of the DDE \cite{breda2006chebyshev}.
Already for a relative small number of Chebyshev nodes the dominant eigenvalues of $\mathbf{M_C}$ yield a good estimate for the dominant eigenvalues of the original DDE \cite{bokharaie2014small} and the systems considered in this paper. \\
Since the matrix $\mathbf{M_C}$ is of dimension $3N(M+1) \times 3N(M+1)$ and a sufficient number of Chebyshev nodes $M$ depends on the considered delay $\tau$ and the system, solving the eigenvalue problem can become computationally expensive.
While the additional information (i.e. eigenvalues and eigenvectors) supplied by the Chebeyshev Collocation method can be useful to gain a deeper inside into the system's dynamics, the critical set of parameters can be more efficiently calculated by using the method presented in Sec.~\ref{sec:theory_stablobes}.

\section{\label{sec:results}Results}
In this section, we discuss the influence of changing inertia, control parameters and time delays on the power grid dynamics and stability.
We consider three different network topologies.
Their parameters will all be chosen from openly available data.
First, we consider a system consisting of two control areas in subsection \ref{subsec:res_twoarea}.
The basic concepts and the general behavior for varying inertia will be discussed.
As an example for a larger network we present results for a Cayley tree network in subsection \ref{subsec:res_bethe}.
Finally, we present results for an network that more closely resembles the control area network of continental Europe in subsection \ref{sec:res_entsoe}.\\ 
In all simulations, we assume that there are no stationary flows between the control areas, which means that the stationary injected power and the stationary power phase angles are zero, i.e., $P_{i,0}=0$ and $ \theta_{i,0}=0$ for $i=1,\ldots,N$. 

\subsection{\label{subsec:res_twoarea}Two Area Network}
The considered network is constructed by separating continental Europe into two control areas. Parameters were chosen to be consistent with the guidelines for load-frequency control released by the ENTSO-E \cite{handbook2004policy, handbook2009policy} and with data provided by the ENTSO-E transparency platform \cite{entsoetrans}. If not stated otherwise, the parameters shown in Table \ref{tab:res_simulation_parameters} were used. The data set describing the sum of all generated power ("Actual Total Load")\cite{entsoetrans} was used to approximate the sum of rated power $S_B$ of the entire continental Europe region.

\begin{table}
    \begin{tabular}{lll}
    \toprule
    Parameter& Symbol &Value \\
    \midrule
    inertia constant & $H$ & 6s \\
    total rated power&$ S_{B,\text{total}}$&  306350.7 MW\\
    rated power area $i$ & $S_{B,i}=S_{B}$ & $S_{B,\text{total}}/N$ \\
    frequency dependant damping & $k_{l,i}$ & $0.01 \frac{1}{\text{Hz}} \cdot S_{B}$\\
    transmission Capacity& $C_{ij}$ & 0.025 $S_{B}$ \\
    total network power freq. characteristic& $\lambda_{\text{total}}$ & $\frac{19000}{2 \pi}$ MW/Hz\\
    network power freq. char. area $i$ & $\lambda$ & $\frac{S_{B,i}}{S_{B,\text{total}}} \cdot \lambda_{\text{total}} $\\
    proportional constant of SC &$K_P$ & 0.4 \\
    integral constant of SC &$K_I$ & 1/120 $s^{-1}$\\
    \bottomrule
    \end{tabular}
    \caption{Standard parameters used in the simulations. If not otherwise indicated, these parameters were used to set up the different systems. The parameter ranges were chosen comparable to the parameters in the European power grid \cite{handbook2004policy, handbook2004policy, handbook2009policy} and data obtained from the ENTSO-E transparency platform \cite{entsoetrans}.}
    \label{tab:res_simulation_parameters}
\end{table}

\subsubsection{\label{subsubsec:twoarea_homogen} Homogeneous Inertia}
\begin{figure}
    \centering
    \includegraphics[width=\columnwidth]{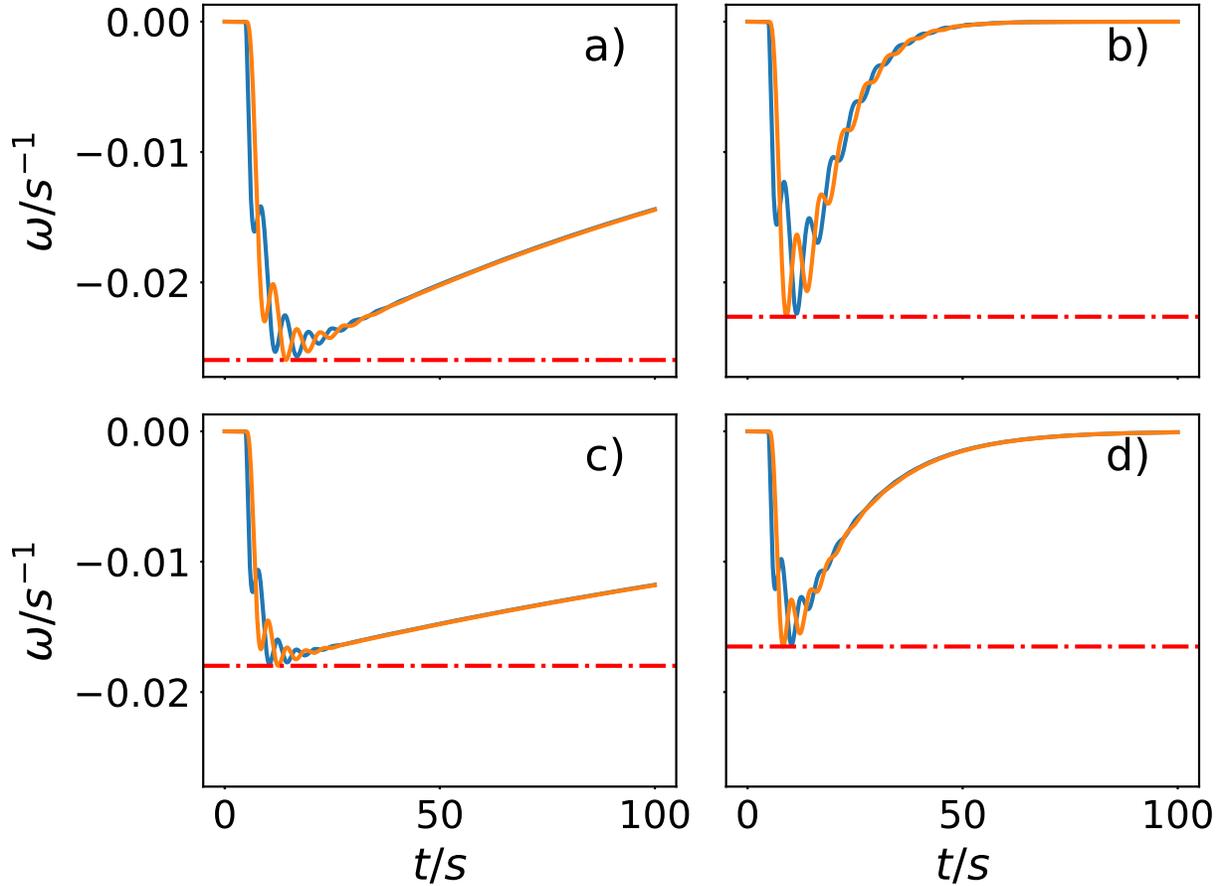}
    \caption{Influence of the gains $K_P$ and $K_I$ of secondary control on the dynamics of a two control area system for $\tau=0$s. The frequency deviation $\omega$ for the two control areas (blue and orange) is shown for different settings of the control gains. $K_P=0.1$ in a) and b) and $K_P=0.7$ in c) and d). $K_I=1/120 \; \text{s}^-1$ in a) and c) and $K_I=1/10 \; \text{s}^-1$ in b) and d). The horizontal red dash-dotted line indicates the lowest frequency deviation $\omega_i$ that occurred due to the disturbance.}
    \label{fig:twoarea_gains}
\end{figure}

\begin{figure}
    \centering
    \includegraphics[width=\columnwidth]{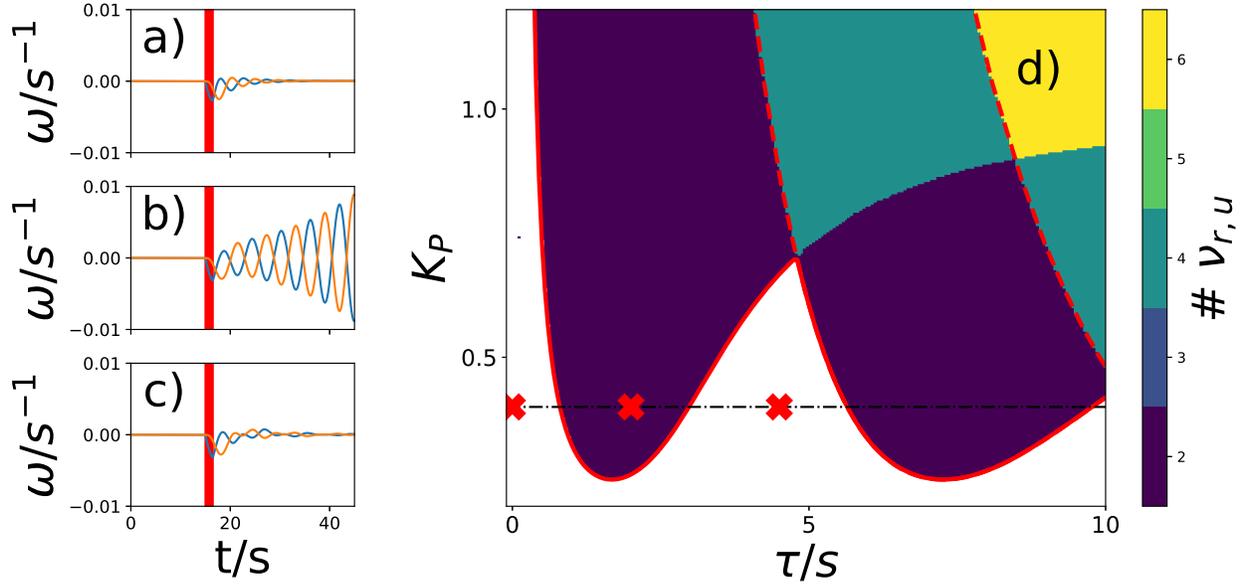}
    \caption{Dependence of the stability of synchronous operation on the delay $\tau$ of secondary control. 
    Left: Time domain simulations for $\tau=0$s (top), $\tau=2$s (middle) and $\tau=4.5$s (bottom).
    Right: Number of unstable roots $\nu_{r,u}$ as a function of proportional gain of secondary control $K_P$ and delay $\tau$.
    The red solid line indicates the stability lobes that separate the regions where the fixed point of synchronous operation is stable and unstable.
    Using Eq.\eqref{eq:periodicity} the stability border can in principle be created by shifting the first lobe, which results in the red dashed line.
    As expected, the red dashed and the red solid line overlap. 
    Parameter combinations for the time domain simulations on the left side are indicated by the three red crosses in the stability chart on the right side.}
    \label{fig:res_twoarea_time_cheb}
\end{figure}

\begin{figure}
    \centering
    \includegraphics[width=\columnwidth]{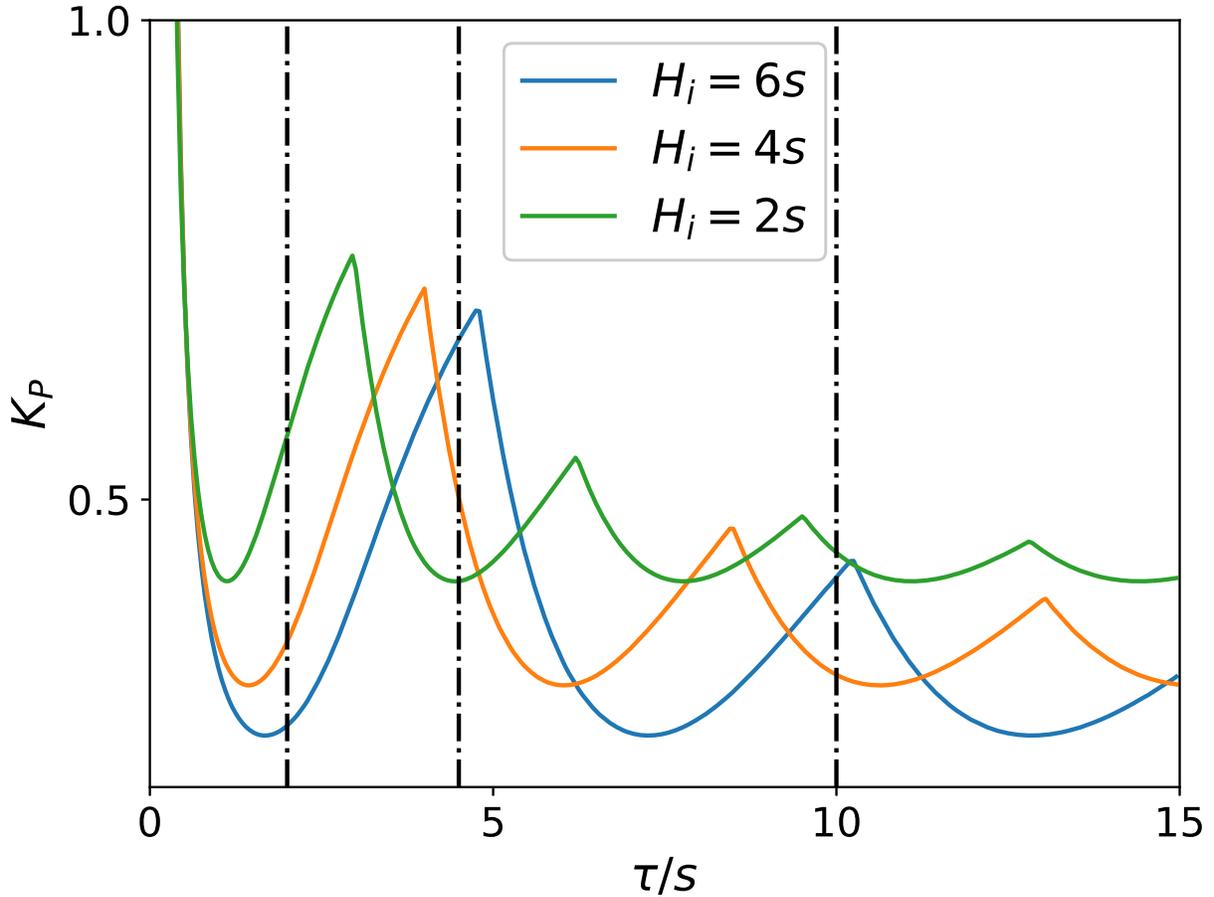}
    \caption{Stability lobes showing the proportional gain of secondary control $K_P$ where the stability behavior of the fixed point changes from stable (below) to unstable (top). Different lines indicate the lobes for different inertia constants $H_i$. Vertical dash-dotted lines correspond to the three delays chosen in Fig.~\ref{fig:res_twoarea_hom_domi}}
    \label{fig:res_twoarea_diffH_hom}
\end{figure}

\begin{figure}
    \centering
    \includegraphics[width=\columnwidth]{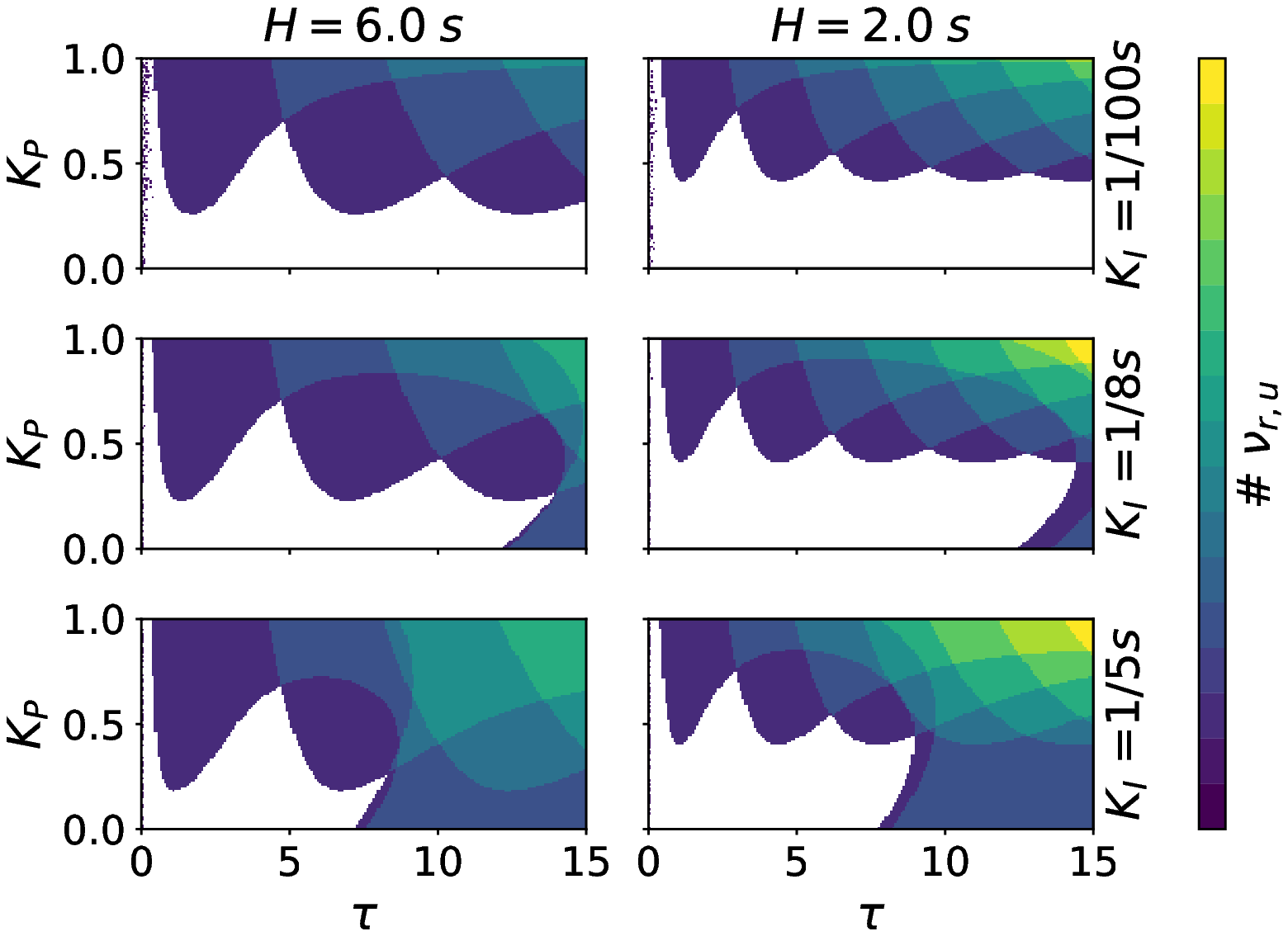}
    \caption{Number of unstable eigenvalues as a function of the proportional gain $K_P$ and delay $\tau$ for the two area network. Two different inertia constants $H_i=6$s (left) and $H_i=2$s (right) and three different integral gains $K_I=1/100$s, $K_I=1/8$s and $K_I=1/5$s (from top to bottom) are used. Larger integral gains (i.e. faster relaxation times $T_N$) decrease the area for stable grid operation (white).}
    \label{fig:res_twoarea_tninfl}
\end{figure}

\begin{figure}
    \centering
    \includegraphics[width=\columnwidth]{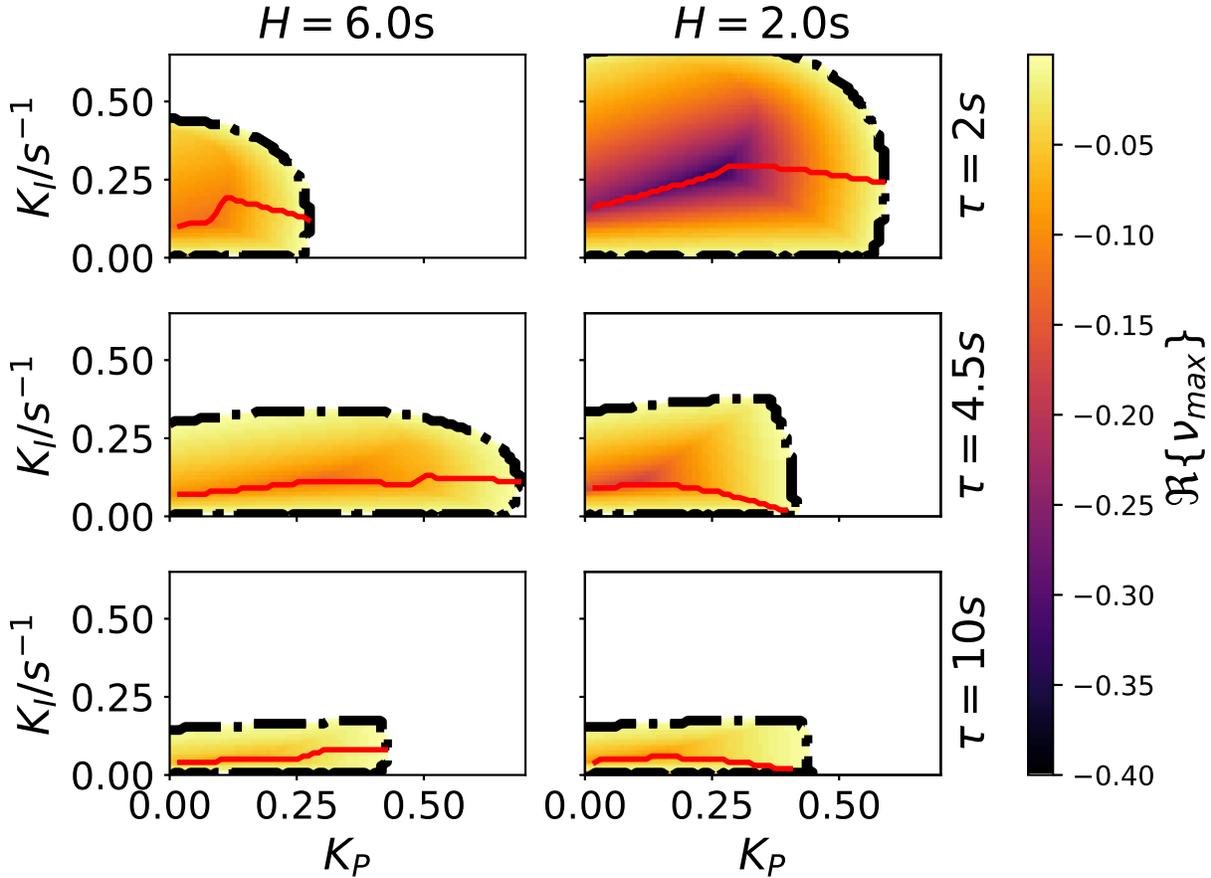}
    \caption{Real part of the dominant eigenvalue $\nu_{max}$ as a function of the proportional gain $K_P$ and integral gain $K_I$ for homogeneously distributed inertia $H_i=H=6$s (left) and $H_i=H=2$s (right). The delay $\tau$ increases from top to bottom. $\nu_{max}$ is only shown in the stable region. The red line indicates the minimal $\nu_{max}$ for a given $K_P$.}
    \label{fig:res_twoarea_hom_domi}
\end{figure}

We are interested in the interplay between the control parameters $K_P, K_I$ and the delay $\tau$ on the stability of the reference state of the power grid. At first, it is helpful to understand the principle influence of the two tunable gains of secondary control on the dynamics. This can be done by solving Eq.~\eqref{eqn:full_eqn} numerically using a solver for delay differential equations \cite{jitcxde}. For the simulation the system was initialized at the fixed point and a disturbance is introduced to one of the two areas. In this case, the disturbance is a sudden increase of load that occurs after a few seconds and persists for the duration of the simulation. In practice, this disturbance could be caused by a large load connecting to the network or the tripping of a line disconnecting a specific generation unit. $P_{\text{PC}, i}$ and $P_{\text{SC},i}$ work in tandem to limit the deviation and restore the pre-disturbance state.\\
The equations were first solved for no delay ($\tau = 0$) and different settings of $K_P$ and $K_I$. The results can be seen in Figure \ref{fig:twoarea_gains}. 
While $K_P$ mainly influences the maximal absolute frequency deviation (or nadir), large $K_I$ results in a faster restoration of the reference value $\omega_0$. $T_N=K_I^{-1}$ can be understood as the time that the system takes to bring the frequency deviation back to zero. It has to be mentioned that tuning $K_P$ and $K_I$ can have different targets (i.e. reducing return time or avoiding overshoot) and is by no means trivial already for the delay-free case ($\tau=0$).\\
Now, we will examine how the control gains, the delay $\tau$ and inertia influence the stability of the fixed point. Time domain simulations of the nonlinear network dynamics for different values of the delay $\tau$ can be seen in Fig.~\ref{fig:res_twoarea_time_cheb} a)-c). Here the disturbance is characterized by an increased load in the interval $t \in [15,16.5]$s (see shaded area in Fig.~\ref{fig:res_twoarea_time_cheb} a)-c)). For the system without delay ($\tau=0$s), the network returns to the synchronous operation at the reference frequency $\omega=0 \; \text{s}^{-1}$. For a delay $\tau=2$s the disturbance increases and the system does not return to the synchronous reference state. Increasing the delay further to $\tau=4.5$s, the fixed point is stable again. This behavior agrees with the results from the Chebyshev collocation method and the identification of the stability boundaries as described in section \ref{sec:theory}. In Fig~\ref{fig:res_twoarea_time_cheb}d) the number of characteristic roots with positive real part derived from the Chebyshev collocation method are shown by the shaded regions. The boundaries between stable and unstable behavior derived from the Chebyshev method fit nicely with the stability lobes (red solid line) derived from the characteristic equation. Indeed, for the chosen $K_P=0.4$ (dash-dotted horizontal line) the stability behavior changes from stable at $\tau=0$s to unstable at $\tau=2$s and stable again at $\tau=4.5$s (red crosses).\\ 
In general, the stability of the fixed point of synchronous operation depends in a complex way on the choice of the tunable gains and on the magnitude of delay $\tau$.
In Fig.~\ref{fig:res_twoarea_time_cheb}d) it can be seen that the number of unstable eigenvalues changes by two, when crossing the border of a stability region.
In this case, a complex conjugate pair of eigenvalues crosses the imaginary axis, thereby changing the number of unstable roots by two, indicating that a Hopf-bifurcation occurs.
When crossing the lobe from the region with zero unstable roots to a region with two unstable roots, the fixed point ceases to be stable and the dynamics evolve into a limit cycle.
Thus, in that case,  the oscillations caused by a small disturbance do not damp out but grow until the dynamics reach the limit cycle behavior.
This persistent oscillatory behavior is not desirable for a power system and might cause severe damage.\\
As more and more inverter-connected generation replaces conventional generators with large rotating masses, the inertia (characterized by $H_i$) decreases.
The effect of a homogeneous change of the inertia on the stability lobes is presented in Fig. \ref{fig:res_twoarea_diffH_hom}.
In general, larger values of $K_P$ corresponding to stable grid operation are possible if the inertia in the system decreases homogeneously.
In addition, in this two area example with homogeneous parameters the peaks in the stability lobes move to lower delays $\tau$ for decreasing inertia constants $H_i$.
This is consistent with results from the literature on machine tool chatter \cite{altintas2004chatter, zatarain2010lobes, otto2014extension}, and an explanation for the observed behavior can be given as follows. Lower inertia constants $H_i$ lead to higher eigenfrequencies, which means that the width of the stability lobes decreases (the distance between two peaks of the stability lobes).
Moreover, lower inertia (and higher eigenfrequencies) leads to a higher damping ratio of the oscillators, and higher damping ratios increase the minimum of the stability lobes.\\ 
As defined above, secondary control has two tunable gains: the proportional gain $K_P$ which gives the reaction to the error measured at $t-\tau$ and the integral gain $K_I$ giving the reaction to the error integrated over the past up to $t-\tau$. In the previously discussed figures, only the proportional gain $K_P$ was varied. $K_I$ was fixed at $K_I=1/120 \; \text{s}^-1$, which is a realistic value for the continental European power grid \cite{handbook2009policy} (cf. Table~\ref{tab:res_simulation_parameters}). The effect of the integral gain $K_I$ on the stability of the reference state can be seen in Fig. \ref{fig:res_twoarea_tninfl}. Faster secondary control (larger $K_I$) leads to a lower parameter range, where a stable reference state can be achieved. In particular, there is a limiting delay $\tau$ which decreases with increasing $K_I$.
For delays larger than this value, which depends also slightly on the proportional gain $K_P$, no stable grid operation is possible.\\ 
In addition to the question if the fixed point is stable or not for the chosen control parameters $K_P$ and $K_I$ over a given range of delays, the optimization of the control parameters with respect to a fast and smooth transition to the pre-disturbance state might be interesting.
As mentioned above, tuning of the parameters of a PI controller is by no means trivial already for the delay-free case. 
Providing a concrete strategy for the tuning in case of a time delay goes beyond the scope of this paper.
However, we would like to present the real part $\nu_\text{max}$ of the dominant characteristic root, which describes the asymptotic exponential behavior of disturbances in the neighborhood of the reference state. 
For $\nu_\text{max}>0$ disturbances grow exponentially and the reference state is unstable. It might be desirable to choose the gains $K_P$ and $K_I$ so that $\nu_\text{max}$ is as negative as possible, ensuring that disturbances decay quickly.
The results for the two area example are shown in Fig.~\ref{fig:res_twoarea_hom_domi}.
The dependence of $\nu_\text{max}$ on $K_P$ and $K_I$ is not monotonic but rather complex.
However, in general, a lower inertia enables more negative $\nu_\text{max}$ ($\min \nu_\text{max}\approx-0.128$s$^{-1}$ for $H=6$s, $\tau=2$s vs. $\min \nu_\text{max}\approx-0.344$s$^{-1}$ for $H=2$s, $\tau=2$s).
Moreover, for increasing time delay $\tau$ the maximum real part $\nu_\text{max}$ of the dominant characteristic roots increases ($\min \nu_\text{max}\approx-0.132$s$^{-1}$ for $H=6$s, $\tau=4.5$s vs. $\min \nu_\text{max}\approx-0.193$s$^{-1}$ for $H=2$s, $\tau=4.5$s). 
Note, that the general behavior of how the system reacts on disturbances depends also on the other characteristic roots and nonlinear effects.\\
\subsubsection{\label{subsubsec:twoarea_inhomo}Inhomogeneous Inertia}
In the previous section, we considered a simplified control area network with homogeneously distributed inertia. As it is unlikely that renewable inverter-connected generation facilities will be equally distributed in the control area network, we consider the case of inhomogeneously distributed inertia. To highlight the effects of homogeneous and heterogeneous distributions of the inertia, we compare two distinct cases: a homogeneous case with the inertia constants in the two areas are set to $H_i=4$s and an inhomogenous or distributed case with the inertia constants chosen as $H_1=2$s and $H_2=6$s.
The total inertia of the two cases is the same.\\ 
\begin{figure}
    \centering
    \includegraphics[width=\columnwidth]{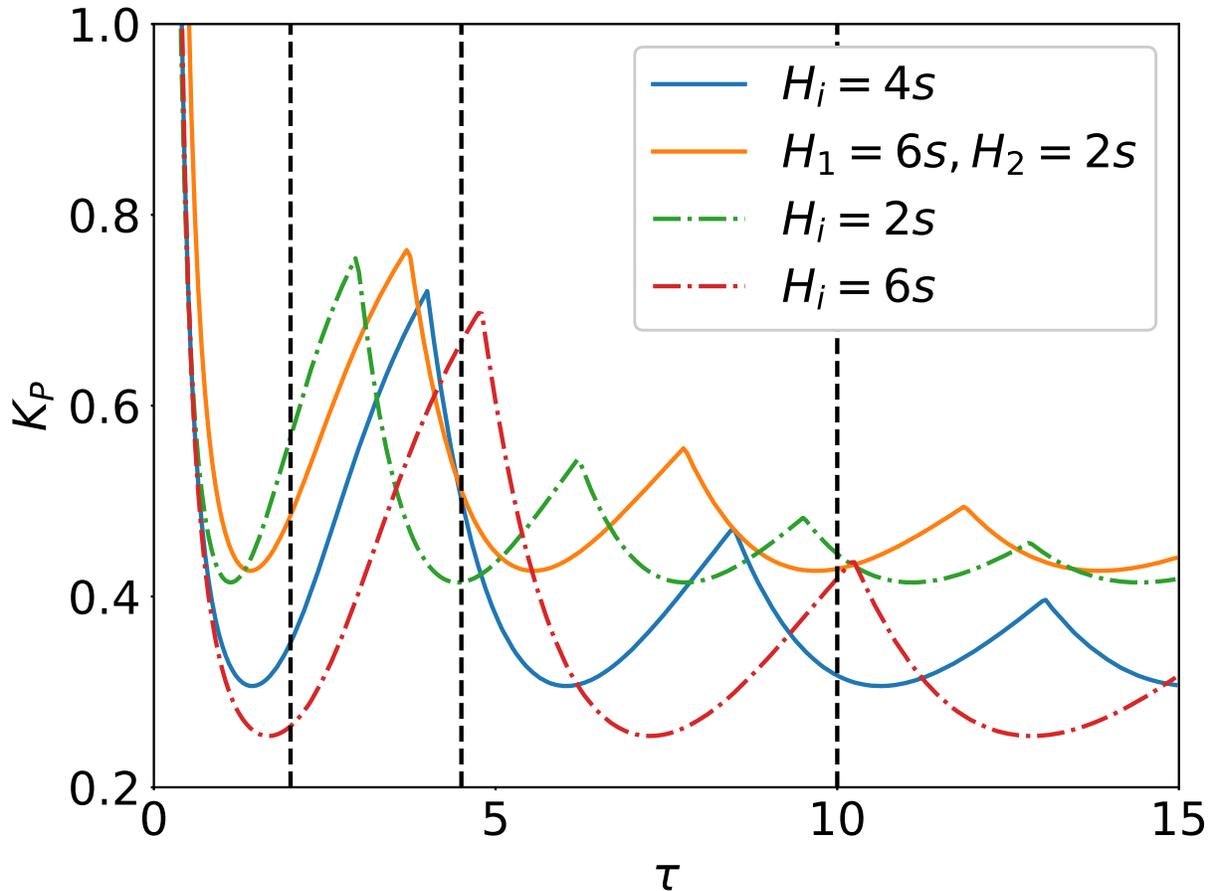}
    \caption{Effect of distributed inertia in the two area example. Solid lines correspond to the stability lobes for a system with homogeneous (blue) and inhomogeneous (orange) inertia and the same total inertia. Dash-dotted lines indicate the stability lobes with homogeneously distributed inertia and inertia constants of the two different inertia constants in the inhomogeneous case. Vertical dashed lines correspond to the delays chosen in Fig.~\ref{fig:res_twoarea_inhom_dom}.}
    \label{fig:res_twoarea_inhom_lobes}
\end{figure}
The resulting stability lobes can be found in Fig.~\ref{fig:res_twoarea_inhom_lobes}. Distributing the inertia inhomogeneously over the two control areas results in a larger stable region in the $K_P$-$\tau$ plane. In particular, the comparison with different stability lobes for homogeneously distributed inertia shows that the stable regions in $K_P$-$\tau$ plane is almost as large as the one for the lowest chosen inertia constant $H_s=2$s.\\
\begin{figure}
    \centering
    \includegraphics[width=\columnwidth]{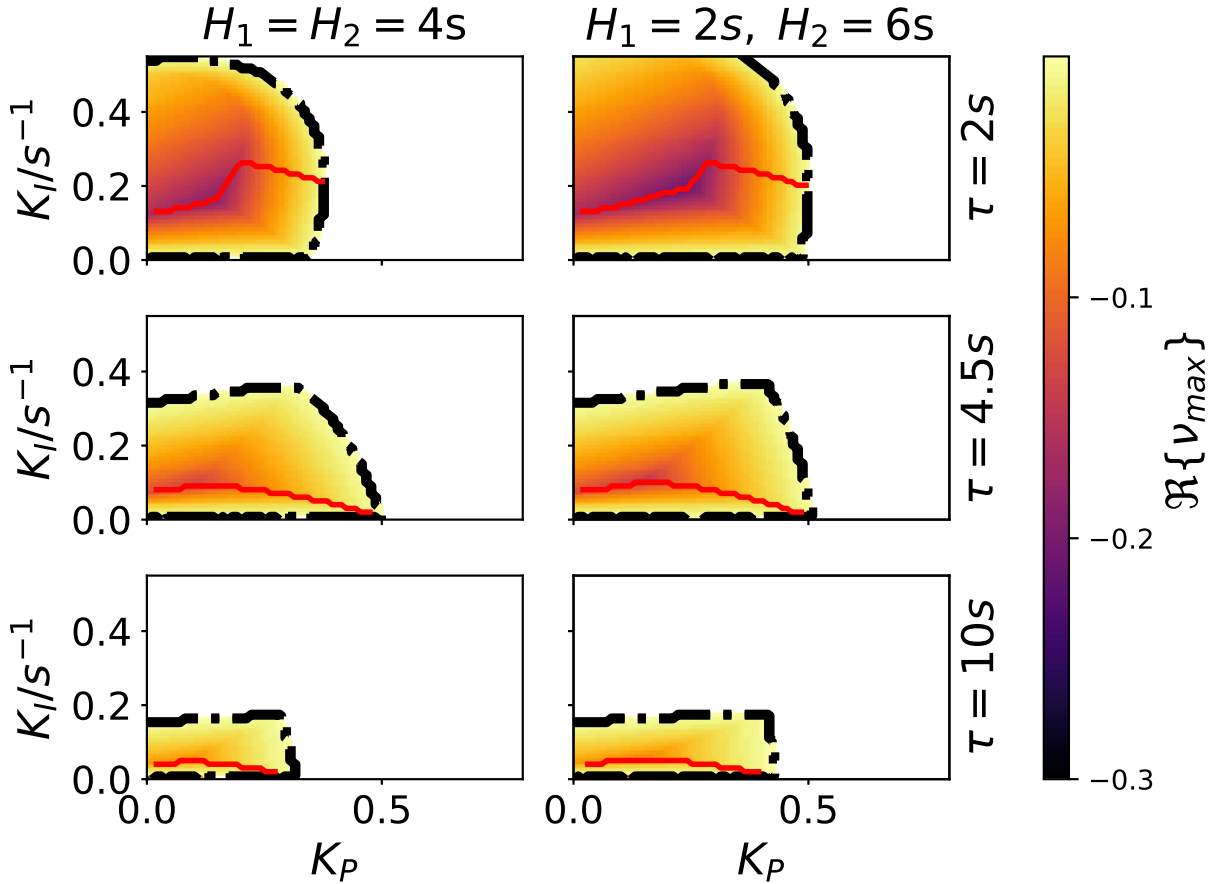}
    \caption{Real part $\nu_{max}$ of the dominant characteristic root as a function of the control gains $K_P$ and $K_I$ for homogeneously (left) and heterogeneously (right) distributed inertia. The results are shown for three different delays $\tau=2$s, $\tau=4.5$s, and $\tau=10$s (from top to bottom), which are marked by vertical dashed lines in Fig.~\ref{fig:res_twoarea_inhom_lobes}. The dash-dotted line indicates the stability boundary for the fixed point.  $\nu_{max}$ is only shown in stable regions. The red line indicates the minimal $\nu_{max}$ for a given $K_P$.}
    \label{fig:res_twoarea_inhom_dom}
\end{figure}
Fig.~\ref{fig:res_twoarea_inhom_dom} shows the stability boundary and the real part of the dominant characteristic root in the parameter plane of the control gains.
One can see that also slightly larger $K_I$ values, corresponding to stable grid operation, are possible for inhomogeneously distributed inertia. 
Especially for $\tau=2$s, a proper tuning of $K_P$ and $K_I$ makes a more negative real part of the dominant eigenvalue possible for the system with inhomogeneously distributed inertia ($\min \nu_\text{max}\approx-0.188$s$^{-1}$ for $H_1=H_2=4$s vs. $\min \nu_\text{max}\approx-0.212$s$^{-1}$ for $H_1=2$s and $H_2=6$s). 
Thus, a faster decay of disturbances can be expected. 

\subsection{\label{subsec:res_bethe} Cayley Tree}
\begin{figure}
    \centering
    \includegraphics[width=\columnwidth]{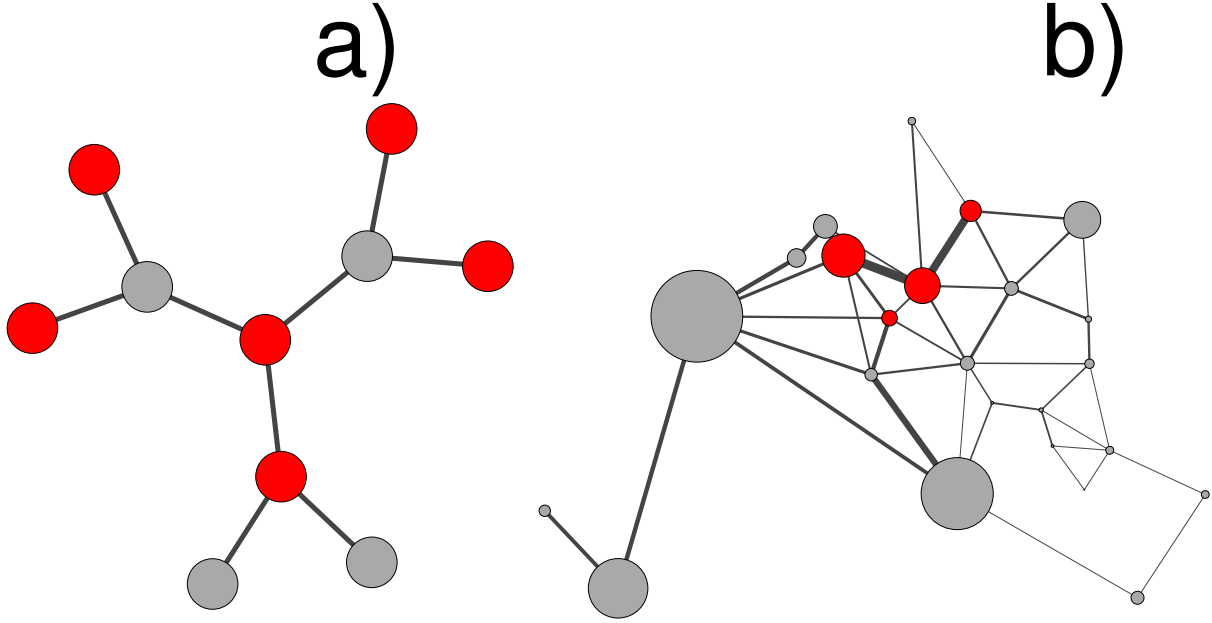}
    \caption{Larger control area networks. a): tree-like network commonly known as Cayley tree.
    Here with a coordination number of three and two layers resulting in 10 total control areas.
    Red color indicates the control areas that have reduced inertia $H_{low}$ in the case of inhomogeneously distributed inertia.
    b): Example of the control area network of continental Europe.
    Red color indicates the 4 german TSOs that have lower inertia in the scenario with inhomogeneously distributed inertia.
    Sizes of vertex and links are proportional to rated power $S_{B,i}$ and transmission capacities $C_{ij}$, respectively.}
    \label{fig:larger_networks}
\end{figure}

In this section, we test if the results of the two area example can be also found in a larger network of control areas.
A tree like topology (Cayley tree) with a total number of $N=10$ control areas (see Figure \ref{fig:larger_networks}) was chosen.
This system will be used a stepping stone, to understand the results from the system derived by data in Sec.~\ref{sec:res_entsoe}.
The remaining setup is similar to the one used for the two area system. 
The total base power $S_{B}$ was distributed to the base power $S_{B,i}=S_{B}/N$ of the individual control areas and the transmission capacities were chosen as $C_{ij} = 0.025 \cdot S_{B,i}$.
The dominant roots and the stability lobes were determined as described in sections \ref{sec:theory_chebyshev} and \ref{sec:theory_stablobes}, respectively.
Similar to Sec.~\ref{subsec:res_twoarea}, homogeneously and inhomogeneously distributed inertia are considered.
In the homogeneous case, the inertia constants for every control area $i$ are chosen as $H_i=4$s. 
For the inhomogeneous case, six areas were chosen for a smaller inertia $H_{low}=8/3$s (red colored control areas in Fig.~\ref{fig:larger_networks}a)).
Inertia constants of the remaining areas were set to $H_{high}=6$s.
This indicates a power system, where the amount of conventional generation in the some regions was replaced by generation by solar panels and wind turbines.
While this choice is somewhat arbitrary, the expansion of renewables will be region specific. 
For example, since there is a larger potential for generation by wind in the northern coastal regions and a higher potential for generation by solar panels in the southern Europe, expansion of renewable is also more likely to occur inhomogeneously and in a fashion specific to the present potentials \cite{zappa2018analysing}.
Again, the total inertia does not change compared to the homogeneous case with $H_i=4\;\text{s} \forall i$.
In summary, the inertia was distributed unevenly throughout the system, yet the transmission capacities $C_{ij}$ and the size in terms of power $S_{B,i}=S_{B}/N$ are constant.\\
\begin{figure}
    \centering
    \includegraphics[width=\columnwidth]{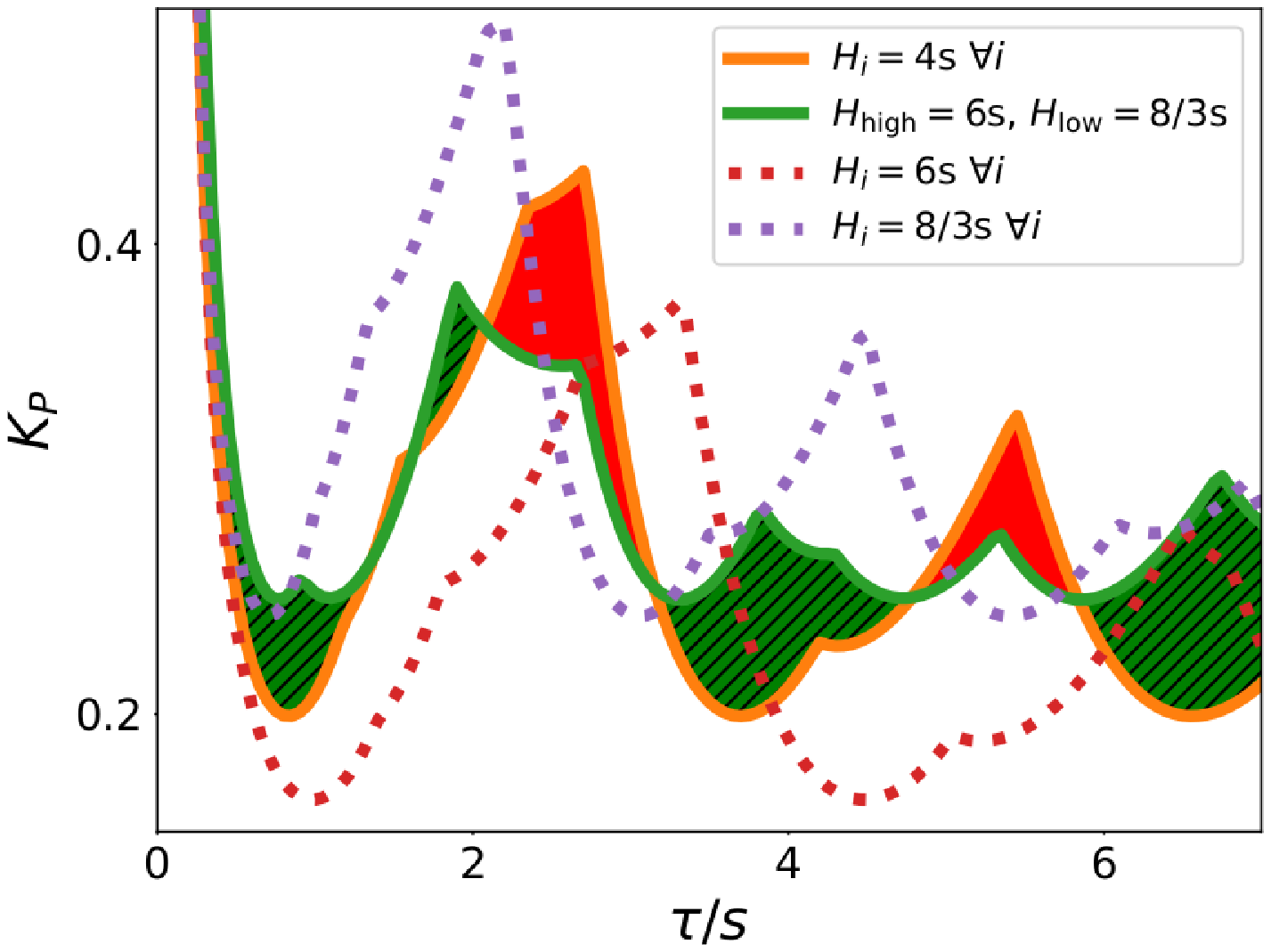}
    \caption{Stability lobes for the Cayley tree with $N=10$ control areas for homogeneously and inhomogeneously distributed inertia.
    The sum of inertia is equal in both cases.
    The integral gain was chosen as $K_I=1/120 \; \text{s}^-1$. Solid lines show the stability lobes i.e. the lowest curves on which an eigenvalue is purely imaginary.
    Dotted lines indicate the lobes for systems with homogeneously distributed inertia.
    For these lobes, the inertia constants are the same ones that can be found in the individual control areas for the inhomogeneous examples.
    Distributed inertia leads to an increased region with a stable fixed point in the $K_P$-$\tau$ plane indicated by the green hatched areas and a decrease for red shaded areas.} 
    \label{fig:res_bethe_lobes}
\end{figure} 
The stability lobes for the Cayley tree are shown in Fig.~\ref{fig:res_bethe_lobes}. 
In the distributed case, some inertia constants are lowered from $H_i=4$s to $H_{low}=8/3$s, while others are increased to $H_{high}=6$s.
Changing the inertia everywhere modifies the stability chart significantly. 
Since the number of relevant modes is a higher than in the two area example, the picture is more complex than the ones for the two area example.\\
In addition to the lobes for the homogeneous and the inhomogeneous case with equal total inertia (solid lines in Figure \ref{fig:res_bethe_lobes}), stability lobes with homogeneously distributed inertia are shown, where the inertia constants are equal to the two different inertia constants in the inhomogeneous case (dotted lines).\\
The resulting stability lobes for both cases in Fig.~\ref{fig:res_bethe_lobes} show that there are benefits (green shaded hatched regions) and detriments (red shaded regions) to the region where the fixed point is stable.
The minimal tolerable $K_P$ for any delay $\tau$ is higher for the case with inhomogeneously distributed inertia.
Similar to the results of the two area example, the stability lobes for the case with inhomogeneously distributed inertia are closest to the stability lobes with homogeneously distributed inertia corresponding to the lower inertia constant of the inhomogeneous case.
This indicates that the benefits, in terms of linear stability of the fixed point, do not necessitate a system with overall low inertia but that a system with redistributed inertia can be similarly beneficial.
Different combinations for choosing high and low inertia areas were examined.
The discussed case in Fig.~\ref{fig:larger_networks}a) was picked to highlight the importance of distributing the inertia intelligently to gain a specific benefit i.e. a higher tolerable $K_P$.
Keep in mind that this is not necessarily also true for the stability border in $K_I$ direction.

\subsection{\label{sec:res_entsoe}Control Area Network of Continental Europe}
\begin{figure}
    \centering
    \includegraphics[width=\columnwidth]{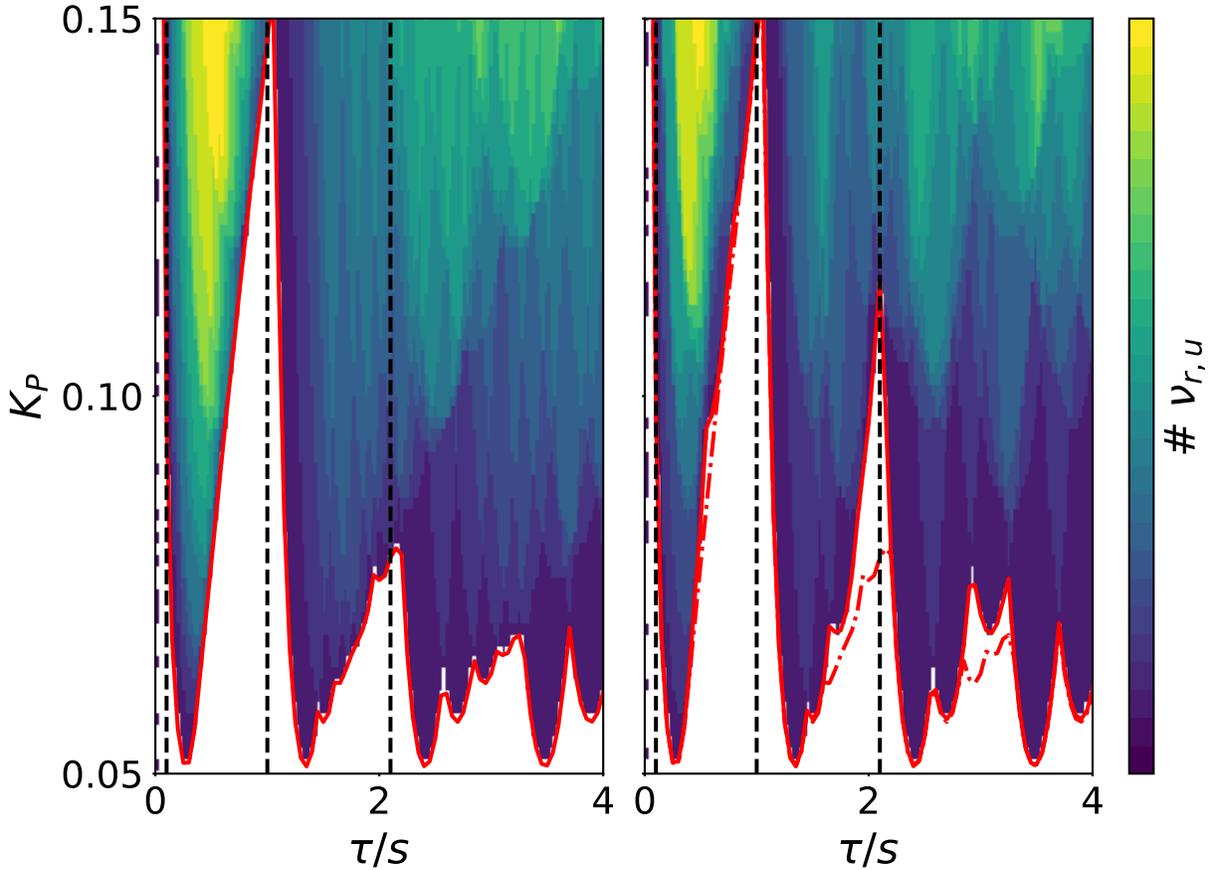}
    \caption{Stability chart for the system extracted from data released by the ENTSO-E with the topology as seen in Fig.~\ref{fig:larger_networks} b). Red lines indicate the stability lobes for the considered system. Left: Homogeneously distributed inertia in each control area with $H_i=6$s. Right: Inhomogeneously distributed inertia $H_{GER}=3$s and all other $H_i=6$s. 
    Red dash-dotted lines are  the stability lobes from the example shown on the left side. The black dashed lines show the delays chosen for Fig.~\ref{fig:res_entsoe_dom}.}
    \label{fig:res_entsoe_cheby}
\end{figure}

\begin{figure}
    \centering
    \includegraphics[width=\columnwidth]{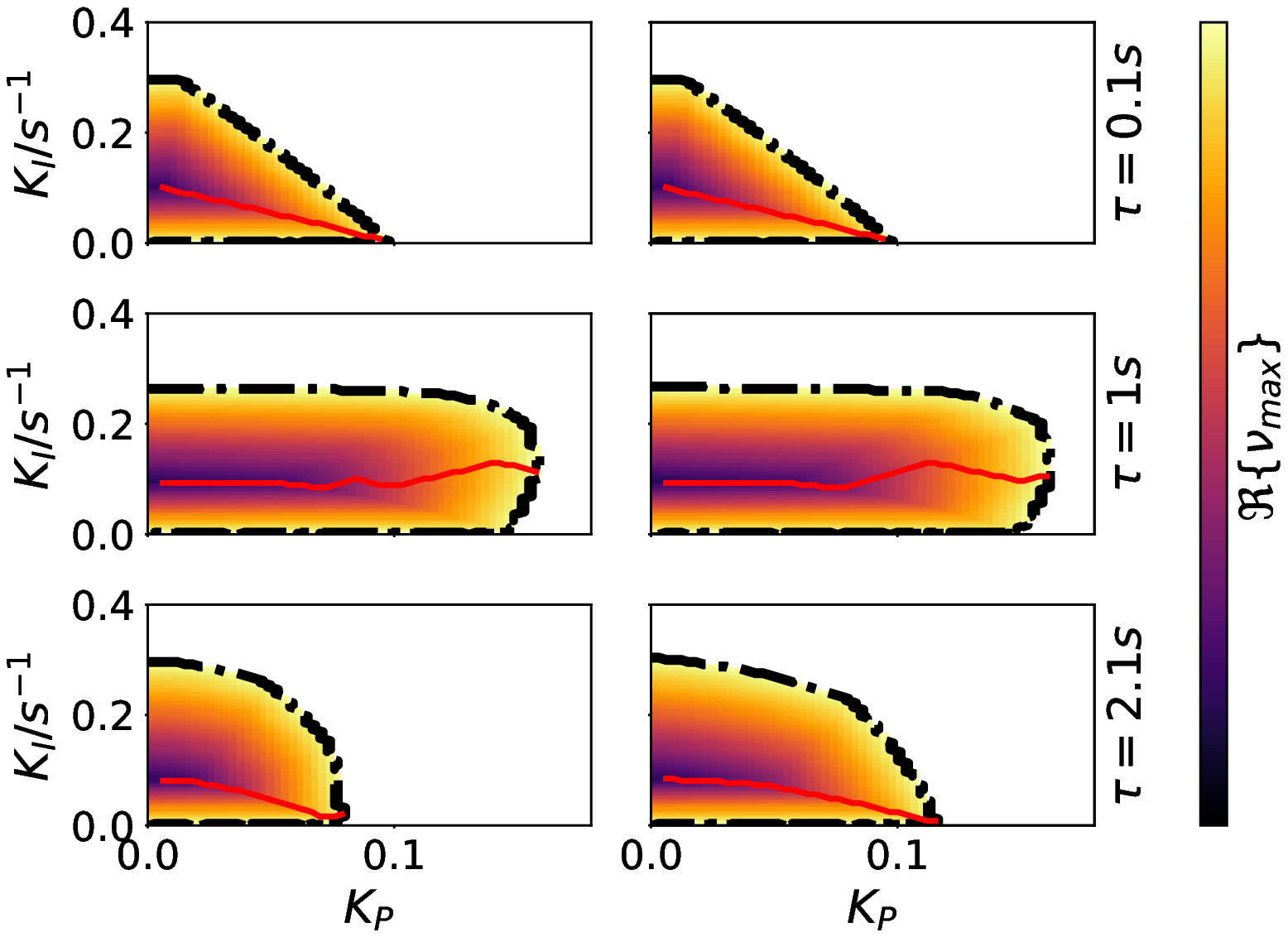}
    \caption{Dominant Eigenvalues for the ENTSO-E control area network.
    The real part of the dominant eigenvalue $\nu_{max}$ as a function of the proportional gain $K_P$ and integral gain $K_I$. Columns show the results for homogeneously distributed (left) and inhomogeneously distributed inertia (right).
    $\nu_{max}$ is only shown in the stable region. A black dash-dotted line separates the stable and unstable regions.
    The red line indicates the minimal $\nu_{max}$ for a given $K_P$.}
    \label{fig:res_entsoe_dom}
\end{figure}

\begin{figure}
    \centering
    \includegraphics[width=\columnwidth]{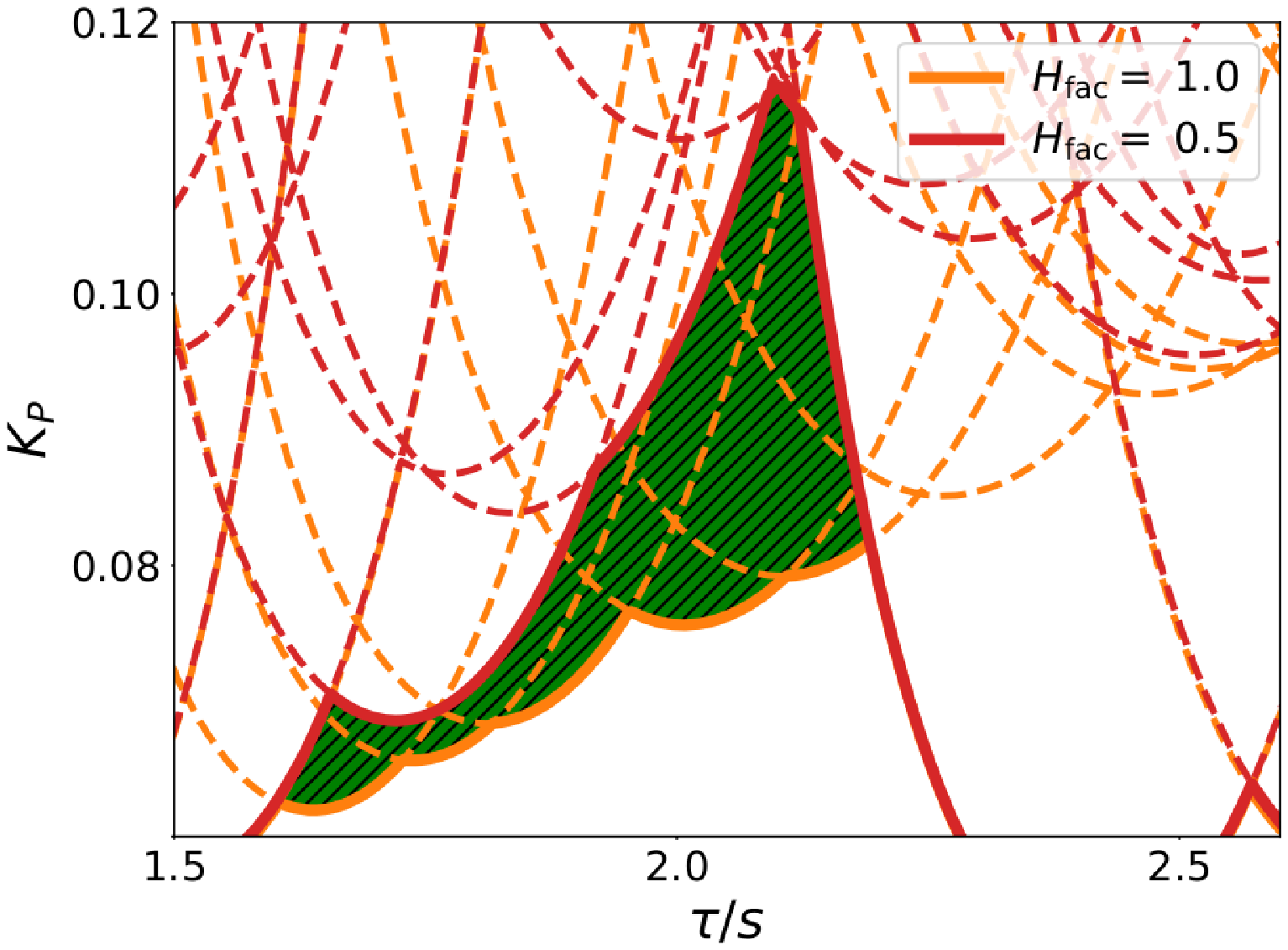}
    \caption{The stability lobes for control area network of Continental Europe are formed by different modes. Lowest curve on which an eigenvalue is purely imaginary (i.e. stability lobes) for the homogeneously distributed inertia with $H_{fac}=1$ (i.e. $H_i=6$s $\forall i$) and the inhomogeneously distributed inertia with $H_{fac}=0.5$ (i.e. $H_{GER}=3$s and $H_i=6$s for all other $i$) are shown by the solid lines.
    The dashed lines show the curves where additional eigenvalues are purely imaginary.
    The benefit (i.e. increase region in $K_P$ with a stable fixed point) is shown by the green hatched area.}
    \label{fig:res_entsoe_multi}
\end{figure}

While some parameters of the previously discussed cases were chosen to be consistent with the transmission system of continental Europe, their topology was simplified.
A more realistic example of the synchronous grid of continental Europe was obtained by analyzing the data provided by the ENTSO-E transparency platform \cite{entsoetrans}.
The values for the size in terms of power $S_{B,i}$ for the individual control areas $i$ were chosen by averaging the daily 'Actual Total Load' in summer for each control area.
The topology of the network in between the individual control areas was determined by analyzing the 'Cross-Border Physical Flow'.
For more details on how this control area network was constructed see Sec.~\ref{app:entsoe} in the appendix.
Since the $n-1$ criteria requires that a maximum of 70\%\cite{kohler2010dena} of the total transmission capacity is used, the maximal recorded flows correspond to 70\% of the available transmission capacity.
The remaining 30\% of backup capacity was evaluated and used as the transmission capacities $C_{ij}$.
The resulting network can be seen in Figure \ref{fig:larger_networks}b).\\ 
In this system, two distinct cases were compared.
One with homogeneously distributed inertia constants $H_i=6$s for every control area and another with inhomogeneously distributed inertia constants $H_i$.
The distributed case was constructed by changing the inertia constants in the four German TSOs  (red colored control areas in Figure \ref{fig:larger_networks}b) by multiplying with a factor $H_{fac} \in \left[0, 1 \right]$ ($H_{GER}=6\text{s} \cdot H_{fac}$) and leaving all other at $H_i=6$s.
Thus in the distributed case, the share of inverter-connected generation to conventional generation was increased in the German TSOs.\\ 
Stability charts for both cases are presented in Fig.~\ref{fig:res_entsoe_cheby}.
The system with inhomogeneously distributed and overall lower inertia constants allows a larger proportional gain $K_P$ that is still in the parameter region for a stable fixed point.
This can be seen especially for intermediate delays (i.e. for $\tau \approx 2$s).
The dominant eigenvalues $\nu_{max}$ for different values of the tunable gains of secondary control are presented in Fig.~\ref{fig:res_entsoe_dom}.
No significant differences between the homogeneous and the inhomogeneous case can be seen for $\tau=0.1$s and $\tau=1$s, whereas for $\tau=2.1$s the stability region increases significantly with decreasing inertia in the German TSOs.\\
In the next step, we varied the inertia constants of the four German TSOs by setting them to $H_{GER}=H_{fac}\cdot 6$s.
\begin{figure}
    \centering
    \includegraphics[width=\columnwidth]{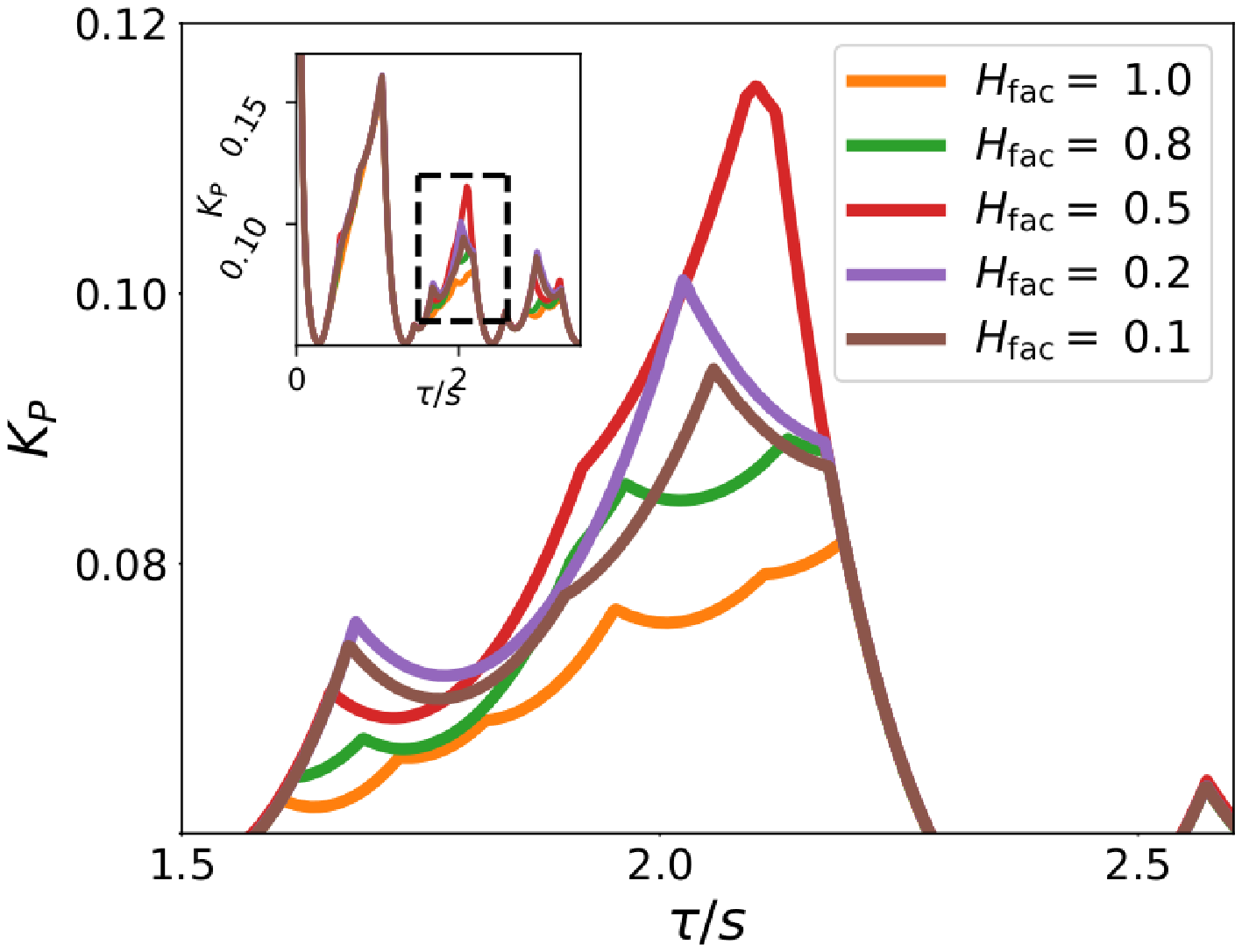}
    \caption{Influence of different levels of inertia in the German TSOs on the stability lobes for the control area network of Continental Europe for intermediate delays ($\tau \in \left[1.5, 2.5\right]$s).
    Inertia change is indicated by $H_{fac}$ giving change in inertia constants for the four German TSOs (i.e. $H_{GER} = H_{fac} \cdot 6s$).
    For example, $H_{fac}=1$ is the case with homogeneously distributed inertia, while for $H_{fac}=0.5$ the inertia in the German TSOs is halved.
    The overall border of stability does not change significantly for every value of the delay $\tau$ as can be seen in the inset plot.
    Intermediate delays, highlighted by the dashed frame in the inset plot, show the largest change regarding the stability border.
    This benefit is largest for $H_{fac}=0.5$.}
    \label{fig:res_entsoe_hfac}
\end{figure}
The results obtained for $H_{fac}=0.5$ in Fig.~\ref{fig:res_entsoe_cheby}, show a change in the stability lobes for delays around $\tau \approx 2$s.
This is also true for different values of $H_{fac}$ as can be seen in Fig.~\ref{fig:res_entsoe_hfac}.
The stability lobes are formed by many different curves, corresponding to different parameter combinations on which eigenvalues are purely imaginary, intersecting with each other.
In the Fig.~\ref{fig:res_entsoe_multi}, the homogeneous case and the case with $H_{fac}=0.5$ are compared.
Changing the inertia constants $H_{GER}$ affects multiple eigenmodes, which shifts the minima of the stability lobes in the $K_P$-$\tau$ plane.
In this example, the inertia was only decreased and not redistributed. 
Still, it is important to be aware that lower inertia not automatically means a larger region in parameter space for which the fixed point is stable.
The lobes presented in Fig.~\ref{fig:res_entsoe_multi} show that reducing the inertia even further by choosing $H_{fac}=0.1$ yields similar results as for $H_{fac}=0.8$, while the benefit for $H_{fac}=0.5$ is the largest.\\
In conclusion, even for the largest network with irregular topology, an decrease of inertia can lead to a larger stability region.
A general statement like: lower inertia increases the region in parameter space where the fixed point is stable can not be made but by knowing the system one can choose where to expand generation by PV and wind to modify the stability chart gaining the benefits as displayed in Fig.~\ref{fig:res_entsoe_cheby} and Fig.~\ref{fig:res_entsoe_hfac}.

\section{\label{sec:conclusion}Summary \& Conclusion}
We employ a model for the frequency dynamics in synchronous control area networks.
Each control area is simulated as one aggregated synchronous machine.
Control mechanisms that are currently used to keep the grid frequency in Europe close to the desired reference frequency, namely primary and secondary control, are included in the model.
A time delay in the feedback control occurs due to data measurement, communication and initiation of a control action. 
Since this is more relevant for the slower secondary control, its reaction to the measured control error was modeled as being delayed by a constant delay $\tau$.\\
Due to the existence of a time delay in the control, the desired reference state of the grid can become unstable. 
Stability lobes separating stable from unstable behavior were found by linearizing the system around the considered fixed point and adapting two existing methods for the stability analysis of DDEs on the power grid model. 
On the one hand, an efficient frequency domain method for the calculation of stability lobes was implemented, and on the other hand, a Chebyshev collocation method was used to approximate the DDE system via a higher dimensional ODE. 
The stability lobes from both methods agree and can be used to select control parameters that ensure stable grid operation. \\
Different network topologies have been examined.
Results obtained by examining a simple two area system, show that the range of values for the tunable gains of secondary control that lead to a stable fixed point increase for lowering the inertia.
This indicates that the expansion of inverter-connected generation (i.e. solar and wind) can be beneficial for the stability of the synchronous state at the reference frequency.
Moreover, distributing the inertia inhomogeneously further increases the region with a stable fixed point. 
Thus, choosing where to install power generation by solar or wind can be advantageous for the system as a whole.
This also holds for larger control area networks, which is shown for the Cayley tree and a system resembling the control area network of continental Europe.\\
The findings suggest that a larger amount of inverter-connected generation can improve the linear stability of desired state of synchronous operation if distributed intelligently.
Therefore, encouraging the development of non-inertia providing renewable generation by subsidies should not only focus on local criteria (e.g. land use) or semi-local (e.g. transmission capacities) criteria but also account for the effects that are the result of the interplay of network topology and delayed control.
Choosing the border between stable and unstable regions explicitly by distributing inertia accordingly throughout the power system might prove useful in guiding the way towards a system highly penetrated by renewable generation.\\
While the presented stability charts give an idea of how the stability lobes are influenced by the different eigenmodes of the power grid model, future work could be related to a deeper understanding of the individual eigenmodes.
This makes designing the stability chart (e.g. Fig.~\ref{fig:res_entsoe_multi}) by shifting individual eigenmodes possible.
Additionally, the presented model can be extended by taking into account, for example, other nonlinearities (e.g. dead band of primary control), more details of the control mechanisms (e.g. simple models for the power dynamics provided by primary and secondary control) and a more realistic delay (e.g. time-dependent by varying between a minimal and maximal delay or distributed by assuming different values for different control areas).


\begin{acknowledgments}
We thank Bruno Schyska, Wilko Heitkötter and Elisavet Proedrou for helpful discussions and proof-reading the manuscript.
PCB acknowledges the funding of the project DYNAMOS by the BMWi (funding code 03ET4027A). 
S. K. acknowledges funding from BMBF CoNDyNet (funding code 03SF0472D) and  CoNDyNetII  (funding code 03EK3055D).
\end{acknowledgments}

\appendix

\section{\label{app:entsoe} Estimating Parameters for the Control Area Network of Continental Europe}
The parameters used in the example of the control area network representing continental Europe were extracted from two data sets from the ENTSO-E transparency platform\cite{entsoetrans}:
\begin{itemize}
    \item 'Actual Total Load': sum of all generation on all grid levels in 15 minutes resolution
    \item 'Cross-Border Physical Flow': flow of electricity from one control area to another control area.
\end{itemize}
The sizes of the control areas, in terms of power, $S_{B,i}$ were estimated by averaging the daily peak in 'Actual Total Load' in summer for each control area $i$. 
Table \ref{tab:para_entsoe} lists the individual control areas with their names, ids and the calculated $S_{B,i}$. \\
The topology and transmission capacities were estimated by analysing the 'Cross-Border Physical Flow'.
This data set provides the flow of electricity between two control areas for every hour.
Assuming that the $n-1$ criteria was obeyed and thus maximally 70\% of the transmission capacities were used, the full capacities $C_{ij, \text{total}}$ were calculated based on the maximal absolute flow.
Only the .99th-quantile of the data points were used to get rid of outliers.
Control areas outside the synchronous grid of continental Europe were ignored.
Additionally, Turkey and Northern Africa were neglected, since the data to calculate the $S_{B,i}$ was missing for these regions.
'Cross-Border Physical Flows' are only recorded if a country border was crossed.
Ergo, the transmission capacities between the four German control areas were determined by using the SciGRID network\cite{SciGRIDv0.2}.
The total transmission capacities $C_{ij, \text{total}}$ between the German control areas were determined by summing the transmission capacities of the gird levels of $110$kV and above of transmission lines that connected the control areas $i$ and $j$.
Table \ref{tab:para_entsoe_links} lists the all links of the control area network consisting of $N=24$ control areas and $45$ links.
A visualization of this network can be seen in Fig.~\ref{fig:larger_networks}b).
The sum of the network power frequency characteristic $\lambda_{\text{total}}=19$ GW/Hz is distributed to the individual control areas according to their share of $S_{B,i}$ giving $\lambda_i = \frac{S_{B,i}}{\sum_i S_{B,i}}\cdot \lambda_{\text{total}}$.\\
This system was used as a basis for the analysis in Sec.~\ref{sec:res_entsoe}.

\begin{table}
    \centering
    \setlength{\tabcolsep}{12pt}

\begin{tabular}{lrr}
\toprule

{Name} &  id &  $S_{B,i}/$MW \\
\midrule
CGES       &   0 &        469.91 \\
Amprion    &   1 &      24857.10 \\
TenneT NL  &   2 &      13687.00 \\
EMS        &   3 &       4558.00 \\
swissgrid  &   4 &       7192.73 \\
TenneT GER &   5 &      20561.20 \\
Energinet  &   6 &       4298.42 \\
ELES       &   7 &       1619.99 \\
PSE SA     &   8 &      21216.20 \\
NOS BiH    &   9 &       1550.22 \\
50Hertz    &  10 &      12225.60 \\
MAVIR      &  11 &       5385.72 \\
CEPS       &  12 &       7940.88 \\
HOPS       &  13 &       2419.00 \\
Elia       &  14 &      10682.20 \\
APG        &  15 &       8070.40 \\
TransnetBW &  16 &       9019.73 \\
Italy      &  17 &      41518.00 \\
RTE        &  18 &      52836.00 \\
SEPS       &  19 &       3503.00 \\
IPTO       &  20 &       7469.00 \\
ESO        &  21 &       4463.00 \\
REN        &  22 &       6530.40 \\
REE        &  23 &      34277.00 \\
\bottomrule
\end{tabular}
    \caption{Name, id and estimated size in terms of power $S_{B,i}$ of the individual control areas for the control area network of continental Europe.}
    \label{tab:para_entsoe}
\end{table}

\begin{table}
\setlength{\tabcolsep}{8pt}
    \centering
\begin{tabular}{rrr}
\toprule
 $i$ &  $j$ &  $C_{ij,total}/$MW \\
\midrule
         0 &          9 &       795.643 \\
         0 &          3 &       567.057 \\
         1 &          4 &      2171.270 \\
         1 &         18 &      3338.320 \\
         1 &          2 &      4461.410 \\
         2 &          5 &      1898.100 \\
         2 &         14 &      4485.110 \\
         3 &          9 &       624.314 \\
         3 &         11 &       695.300 \\
         3 &         13 &       687.143 \\
         3 &         21 &       776.829 \\
         4 &         17 &      6658.570 \\
         4 &         15 &      2406.230 \\
         4 &         18 &      3370.890 \\
         4 &         16 &      4436.180 \\
         5 &         15 &      2431.970 \\
         5 &         12 &      2244.540 \\
         5 &          6 &      2168.680 \\
         6 &         10 &       857.200 \\
         7 &         15 &      1626.070 \\
         7 &         17 &      1754.300 \\
         7 &         13 &      2025.710 \\
         8 &         12 &      2537.200 \\
\bottomrule
\end{tabular}
\hspace*{0cm}
\begin{tabular}{rrr}
\toprule
 $i$ &  $j$ &  $C_{ij,total}/$MW \\
\midrule
         8 &         19 &      1453.710 \\
         8 &         10 &      2693.620 \\
         9 &         13 &      1948.570 \\
        10 &         12 &      2517.430 \\
        11 &         15 &      1572.400 \\
        11 &         13 &      1696.110 \\
        11 &         19 &      2492.210 \\
        12 &         15 &      3281.860 \\
        12 &         19 &      2930.710 \\
        14 &         18 &      4607.560 \\
        15 &         17 &       407.714 \\
        15 &         16 &      2004.700 \\
        16 &         18 &      2421.320 \\
        17 &         18 &      4020.000 \\
        17 &         20 &       731.429 \\
        18 &         23 &      4684.360 \\
        20 &         21 &       775.714 \\
        22 &         23 &      4105.890 \\
         1 &         16 &      3042.000 \\
         5 &         16 &      1976.000 \\
         5 &         10 &      8398.000 \\
         1 &          5 &      9672.000 \\
         & & \\
\bottomrule
\end{tabular}
    \caption{List with the estimated total transmission capacities $C_{ij, \text{total}}$ between the control areas $i$ and $j$ in the example of the control area network of continental Europe discussed in Sec.~\ref{sec:res_entsoe}. The identifying source and target ids are shown in Tab.~\ref{tab:para_entsoe}.}
    \label{tab:para_entsoe_links}
\end{table}

\FloatBarrier

%

\end{document}